\documentstyle[12pt,epsfig,amsmath,amssymb,pstricks]{article}
\begin{document}

\title{\bf If physics is an information science, what is an observer?}

\author{{Chris Fields}\\ \\
{\it 21 Rue des Lavandi\`eres}\\
{\it Caunes Minervois, 11160 France}\\ \\
{fieldsres@gmail.com}}
\maketitle

\abstract{Interpretations of quantum theory have traditionally assumed a ``Galilean'' observer, a bare ``point of view'' implemented physically by a quantum system.  This paper investigates the consequences of replacing such an informationally-impoverished observer with an observer that satisfies the requirements of classical automata theory, i.e. an observer that encodes sufficient prior information to identify the system being observed and recognize its acceptable states.  It shows that with reasonable assumptions about the physical dynamics of information channels, the observations recorded by such an observer will display the typical characteristics predicted by quantum theory, without requiring any specific assumptions about the observer's physical implementation.}



\begin{center}
``\textit{Information}?  \textit{Whose} information?  Information about \textit{what}?''
\end{center}
\begin{flushright}
J. S. Bell (\cite{bell90} p, 34; emphasis in original)
\end{flushright}

\section{Introduction}

Despite over 80 years of predictive success (reviewed in \cite{schloss06}), the physical interpretation of quantum states, and hence of quantum theory itself remains mysterious (for recent reviews see \cite{schloss07, landsman07, wallace08}).  Informally speaking, this mysteriousness results from the apparent dependence of the physical dynamics on the act of observation.  Consider Schr\"odinger's cat: the situation is paradoxical because the observer's act of opening the box and looking inside appears to \textit{cause} the quantum state of the cat to ``collapse'' from the distinctly non-classical superposition $|cat\rangle = \frac{1}{\sqrt{2}}(|alive\rangle + |dead\rangle)$ to one of the two classical eigenstates $|alive\rangle$ or $|dead\rangle$.  The introduction of decoherence theory in the 1970s and 80s \cite{zeh70, zeh73, zurek81, zurek82, joos-zeh85} transferred this mysterious apparently-causal effect on quantum states from what the observer looks at - the system of interest - to what the observer ignores: the system's environment (reviewed by \cite{joos-zeh03, zurek03rev, schloss04}; see also \cite{schloss07, landsman07, wallace08} for treatments of decoherence in a more general context and \cite{bacci07} for a less formal, more philosophical perspective).  Schr\"odinger's poor cat, for example, interacts constantly with the environment within the box - stray photons, bits of dust, etc. - and via the walls of the box with the thermal environment outside.  Components of $|cat\rangle$ thereby become entangled with components of the environmental state $|env\rangle$, a state that spreads at the speed of light to encompass all the degrees of freedom of the entire universe (other than the cat's) as the elapsed time $t \rightarrow \infty$.  To an observer who does not look at the environment, this entanglement is invisible; the components of the environment can therefore be ``traced out'' of the joint quantum state $|cat \otimes env\rangle$ to produce an ensemble of non-interfering, effectively classical states of just the cat, each with a well-defined probability.  Such reasoning about what observers do not look at is employed to derive effectively classical states of systems of interest throughout the applied quantum mechanics literature.  For example, Martineau introduces decoherence calculations intended to explain why the Cosmic Background Radiation displays only classical fluctuations with the remarks: ``Decoherence is, after all, an observer dependent effect - an observer who could monitor every degree of freedom in the universe wouldn't expect to see any decoherence.  However, our goal is to determine a lower bound on the amount of decoherence as measured by any observer ... we trace out only those modes which we must ... and take our system to be composed of the rest'' (\cite{martineau06} p. 5821).  Noting that the setting for these calculations is the inflationary period immediately following the Big Bang, one might ask, ``\textit{Observer}?  \textit{What} observer?  Looking at \textit{what}?''

Ordinary observers in ordinary laboratories interact with ordinary, macroscopic apparatus in order to gain classical information in the form of macroscopically and stably recordable experimental outcomes.  The reconceptualization of physics as an information science that developed in the last quarter of the $20^{\mathrm{th}}$ century, motivated by Feynman's speculation that all of physics could be simulated with a quantum computer \cite{feynman82}, Wheeler's ``it from bit'' proposal that ``all things physical ... must in the end submit to an information-theoretic description'' (\cite{wheeler92} p. 349), Deutsch's proof of the universality of the quantum Turing machine (QTM \cite{deutsch85}) and Rovelli's explicitly information-theoretic derivation of relational quantum mechanics \cite{rovelli96}, reformulated the problem of describing measurement as the problem of describing how observers could obtain classical information in a world correctly described by the quantum mechanical formalism.  Theoretical responses to this reconceptualization can be divided into two broad categories by whether they maintain the standard Dirac - von Neumann Hilbert-space formalism as fundamental to quantum mechanics and adopt information-theoretic language to its interpretation, or adopt information-theoretic postulates as fundamental and attempt to derive the Hilbert-space formalism from them.  Responses in the first category treat decoherence as a fundamental physical process and derive an account of measurement from it; examples include traditional relative-state (i.e. many-worlds or many-minds) interpretations \cite{joos-zeh03, tegmark98, zeh00, tegmark10, wallace10, susskind11}, the consistent histories formulation \cite{griffiths02, hartle08, griffiths11} and quantum Darwinism \cite{zurek03rev, zurek04, zurek05, zurek06, zurek07grand, zurek09rev}.  Those in the second treat measurement as a fundamental physical process; they are distinguished by whether they treat information and hence probabilities as objective \cite{clifton03, bub04, lee11} or subjective \cite{rovelli96, fuchs02, fuchs10, chiribella11, rau11, spekkens11}.

While observers appear as nominal recipients of information in all interpretative approaches to quantum theory, the \textit{physical structure} of an observer is rarely addressed.  Zurek \cite{zurek03rev}, for example, remarks that observers differ from apparatus in their ability to ``readily consult the content of their memory'' (p. 759), but nowhere specifies either what memory contents are consulted or what memory contents might be required, stating that ``the observer's mind (that verifies, finds out, etc.) constitutes a primitive notion which is prior to that of scientific reality'' (p. 363-364).  Hartle \cite{hartle08} characterizes observers as ``information gathering and utilizing systems (IGUSes)'' but places no formal constraints on the structure of an IGUS and emphasizes that the information gathered by IGUSes is ``a feature of the universe independent of human cognition or decision'' (p. 983).  Rovelli \cite{rovelli96} insists that ``The observer can be any physical system having a definite state of motion'' (p.  1641).  Schlosshauer \cite{schloss07} adopts the assumption that appears most commonly throughout the literature: ``We simply treat the observer as a quantum system interacting with the observed system'' (p. 361).  Fuchs \cite{fuchs10} treats observers as Bayesian agents, and not only rejects but lampoons the idea that the physical implementation of the observer could be theoretically important: ``would one ever imagine that the notion of an agent, the user of the theory, could be derived out of its conceptual apparatus?'' (p. 8).  While such neglect (or dismissal) of the structure of the observer is both traditional and \textit{prima facie} consistent with the goal of building a fully-general, observer-independent physics, it seems surprising in a theoretical context motivated by ``it from bit'' and the conceptualization of physical dynamics as quantum computing.

It is the contention of the present paper that the physical structure of the observer is important to quantum theory, and in particular that the information employed by the observer to \textit{identify} the system of interest as an information source must be taken into account in the description of measurement.  This contention is motivated by the intuition expressed by Rovelli, that ``the unease (in the interpretation of quantum theory) may derive from the use of a concept which is inappropriate to describe the world at the quantum level'' (\cite{rovelli96} p. 1638).  On the basis of this intuition, Rovelli rejects the assumption of observer-independent quantum states, an assumption also rejected by quantum Bayesians \cite{fuchs02, fuchs10, rau11, spekkens11}.  The present paper rejects an equally-deep assumption: the assumption of a ``Galilean'' observer, an observer that is simply ``a quantum system interacting with the observed system'' without further information-theoretic constraints.  As the analysis of Rovelli \cite{rovelli96} demonstrates, measurement interactions between a Galilean observer and a physical system can be described in terms of Shannon information, but this can only be done from the perspective of a second observer or a theorist who stipulates what is to count as ``observer'' and ``system.''  The use of Galilean observers in an information-theoretic formulation of physical theory thus requires that the identities of ``systems'' be given in advance.  That this requirement is problematic has been noted by Zurek, who states that ``a compelling explanation of what the systems are - how to define them given, say, the overall Hamiltonian in some suitably large Hilbert space - would undoubtedly be most useful'' (\cite{zurek98rough} p. 1818), and requires as ``axiom(o)'' of quantum mechanics that ``(quantum) systems exist'' (\cite{zurek03rev} p. 746; \cite{zurek07grand} p. 3; \cite{zurek05env} p. 2) as objective entities.  Zurek adopts Wheeler's \cite{wheeler75} view that the universe itself can be considered to be the ``second observer,'' and proposes from this ``environment as witness'' perspective that decoherence provides the physical mechanism by which systems ``emerge'' into objectivity \cite{zurek03rev, zurek04, zurek05, zurek06, zurek07grand, zurek09rev}.  Decoherence is similarly proposed to be the mechanism by which quantum information becomes classical \cite{griffiths07} and by which both Everett branches \cite{tegmark10, wallace10} and the frameworks defining consistent histories \cite{griffiths02, hartle08, griffiths11} are distinguished.  By rejecting the assumption of Galilean observers, the present paper also rejects the idea that the objective existence of systems can be taken as given \textit{a priori}, either by an axiom or by a physical process of emergence.  Instead, it proposes that not just quantum states but systems themselves are definable only relative to observers, and in particular, that quantum systems are defined only relative to classical information encoded by observers.  An alternative approach to understanding quantum theory in informational terms is proposed, one that explicitly recognizes the requirement that observers encode sufficient information to enable the identification and hence the definition of the systems being observed.

That ordinary observers in ordinary laboratories must be in possession of information sufficient to identify systems of interest as classical information sources, not just instantaneously but over extended time, is uncontroversial in practice.  It follows immediately, moreover, from Moore's 1956 proof that no finite sequence of observations of the outputs generated by a finite automaton in response to given inputs could identify the automaton being observed (\cite{moore56} Theorem 2; cf. \cite{ashby56} Ch. 6).  Hence ordinary observers are not Galilean.  The information employed by an ordinary, non-Galilean observer to identify a system being observed is ``pragmatic'' information in the sense defined by Roederer \cite{roederer05, roederer11}, although as will be seen below, without Roederer's restriction of such information to living (i.e. evolved self-reproducing) systems.  That observers must encode such pragmatic information in their physical structures follows from the physicalist assumption - the complement of ``it from bit'' - that all information is physically encoded \cite{landauer99}.  The notion of an ``observer'' as a physical device encoding input-string parsers or more general input-pattern recognizers that fully specify its observational capabilities underlies not only the design and implementation of programming languages and other formal-language manipulation tools (e.g. \cite{kleene67, tan76, hopcroft79}), but also computational linguistics and the cognitive neuroscience of perception (e.g. \cite{marr82, dretske83, rock83, kosslyn94, ullman96, pinker97}).  

It is shown in what follows that when the pragmatic information encoded by ordinary observers is explicitly taken into account, distinctive features of the quantum world including the contexuality of observations, the violation of Bell's inequality and the requirement for complex amplitudes to describe quantum states follow naturally from simple physical assumptions.  The next section, ``Interaction and System Identification'' contrasts the description of measurement as physical interaction with its description as a process of information transfer, and shows how the problem of system identification arises in the latter context.  The third section, ``Informational Requirements for System Identification'' formalizes the minimal information that an observer must encode in order to identify a macroscopic system - a canonical measurement apparatus - that reports the pointer values of two non-commuting observables.  It then defines a \textit{minimal observer} in information-theoretic terms as a virtual machine encoding this minimal required information within a control structure capable of making observations and recording their results.  The following section, ``Physical Interpretation of Non-commutative POVMs'' considers the physical implementation of a minimal observer in interaction with a physical channel.  It shows that if the physical dynamics of the information channel are time-symmetric, deterministic, and satisfy assumptions of decompositional equivalence and counterfactual definiteness, any minimal observer encoding POVMs that jointly measure physical action will observe operator non-commutativity independently of any further assumptions about the observed system.  The fifth section, ``Physical Interpretation of Bell's Theorem, the Born Rule and Decoherence'' shows that the familiar phenomenology of quantum measurement follows from the assumptions of minimal observers and channel dynamics that are time-symmetric, deterministic, and satisfy decompositional equivalence and counterfactual definiteness.  It shows, in particular, that decoherence can be understood as a consequence of hysteresis in quantum information channels, and that the use of complex Hilbert spaces to represent observable states of quantum systems is required by this hysteresis.  The sixth section, ``Adding Minimal Observers to the Interpretation of Quantum Theory'' reviews the ontology that naturally follows from the assumption of minimal observers, an ontology that is realist about the physical world but virtualist about ``systems'' smaller than the universe as a whole.  It shows that any interpretative framework that treats ``systems'' as objective implicitly assumes that information is free, i.e. implicitly assumes that the world is classical.  The paper concludes by suggesting that the interpretative problem of interest is that of understanding the conditions under which a given physical dynamics implements a given virtual machine, i.e. the problem of understanding the ``emergence'' not of ``classicality'' but of observers.

\section{Interaction and System Identification}

The extraordinary empirical success of quantum theory suggests strongly that quantum theory is the correct description of the physical world, and that classical physics is an approximation that, at best, describes the appearance of the physical world under certain circumstances.  Landsman \cite{landsman07} calls the straightforward acceptance of this suggestion ``stance 1'' and contrasts it with the competing view (``stance 2'') that quantum theory is itself an approximation of some deeper theory in which the world remains classical after all.  This paper assumes the correctness of quantum theory; Landsman's ``stance 1'' is thus adopted.  In particular, it assumes \textit{minimal} quantum theory, in which the universe as a whole undergoes deterministic, unitary time evolution described by a Schr{\"o}dinger equation.  The question that is addressed is how the formal structure of minimal quantum theory can be understood physically, as a description of the conditions under which observers can obtain classical information about the evolving states of quantum systems.

As emphasized by Rovelli \cite{rovelli96}, minimal quantum theory treats all systems, including observers, in a single uniform way.  The interaction between an observer and a system being observed can, therefore, be represented as in Fig. 1a: both observer and observed system are collections of physical degrees of freedom that are embedded in and interact with the much larger collection of physical degrees of freedom - the ``environment'' - that composes the rest of the universe.  The present paper adopts a realist stance about these physical degrees of freedom; they can be considered to be the quantum degrees of freedom of the most elementary objects with which the theory is concerned.  The observer - system interaction is described by a Hamiltonian $\mathcal{H}_{\mathbf{O-S}}$; this Hamiltonian is well-defined to the extent that the boundaries separating the observer and the system from the rest of the universe are well-defined.  In practice, however, neither the system - environment nor the observer - environment boundaries are determined experimentally.  The degrees of freedom composing the system $\mathbf{S}$ are typically specified by specifying a set $\lbrace |s_{i}\rangle \rbrace$ of orthonormal basis vectors, e.g. by saying ``let $|S\rangle = \sum_{i} \lambda_{i} |s_{i}\rangle$.''  The set $\lbrace |s_{i}\rangle \rbrace$ is a subset of a set of basis vectors spanning the Hilbert space $\mathbf{H}_{\mathbf{U}}$ of the universe as a whole; it defines a subspace of $\mathbf{H}_{\mathbf{U}}$ with finite dimension $d$ that represents $\mathbf{S}$.  The state of $\mathbf{O}$, on the other hand, is typically left unspecified, and the $\mathbf{O - S}$ interaction is represented not as a Hamiltonian but as a measurement that yields classical information.  Traditionally, measurements are represented as orthonormal projections along allowed basis vectors of the system (e.g. \cite{vonNeumann32}); distinct real ``pointer values'' representing distinct observable outcomes are associated with each of these projections.  In current practice, the requirement of orthogonality is generally dropped and measurements are represented as positive operator-valued measures (POVMs), sets of positive semi-definite operators $\lbrace \mathcal{E}_{\mathit{j}} \rbrace$ that sum to the identity operator on the Hilbert space of $\mathbf{S}$ (e.g. \cite{nielsen-chaung00} Ch. 2).  As shown by Fuchs \cite{fuchs02}, a ``maximally informative'' POVM can be constructed from a set of $d^{2}$ projections $\lbrace \Pi_{j} \rbrace$ on the Hilbert space spanned by $\lbrace |s_{i}\rangle \rbrace$.  The first $d$ components of such a POVM are the orthogonal projections $|s_{i}\rangle\langle s_{i}|$; ``pointer values'' can be associated with these $d$ orthogonal components in the usual way.

\psset{xunit=1cm,yunit=1cm}
\begin{pspicture}(0,0)(16,12)
\put(1,11){(a)}

\put(2.6,9){``Observer''}
\pscurve(2.2,9)(2.1,10.2)(3.2,10.2)
\pscurve(3.2,10)(4.2,10.8)(4.8,9.8)
\pscurve(4.6,9.8)(5.6,8.6)(4.2,8.2)
\pscurve(2.4,9)(2.2,8)(3.4,8.4)
\pscurve(3,8)(4,7.2)(4.6,8)

\put(6.8,10.5){``Environment''}
\psline{<->}(6,9)(10,9)
\put(7.5,9.2){$\mathcal{H}_{\mathbf{O-S}}$}

\put(11.6,9){``System''}
\pscurve(11.3,9.8)(10.2,9)(11.8,8.2)
\pscurve(11.1,9.9)(11.6,11)(12.8,10.4)
\pscurve(12.8,10.6)(14.2,10.6)(14,9.4)
\pscurve(14.2,9.4)(15,8)(13.4,7.8)
\pscurve(13.4,8.2)(12.2,7.2)(11.6,8)

\put(1,6.5){(b)}

\put(3,4.7){Observer}
\pspolygon(2,3.5)(2,6)(6,6)(6,3.5)

\put(6.5,6.2){Information channel}
\put(7,5.2){Intervention}
\put(6.5,5){\vector(1,0){3.5}}
\put(10,4.5){\vector(-1,0){3.5}}
\put(7.3,4){Outcome}

\put(11.8,4.7){System}
\pspolygon(10.5,3.5)(10.5,6)(14.5,6)(14.5,3.5)

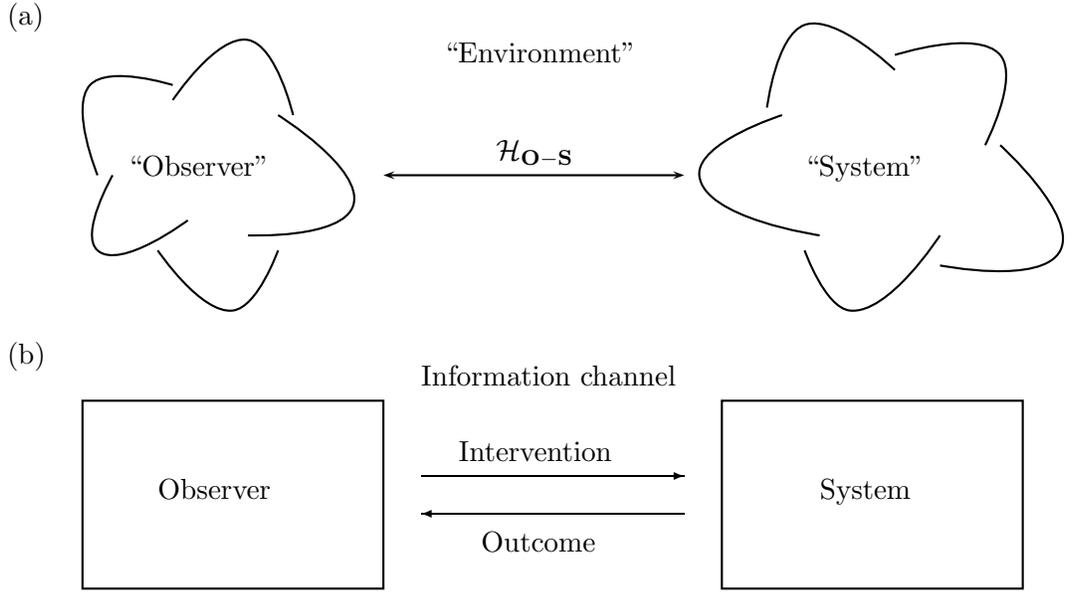
\put(0.5,2.5){\textit{Fig. 1: (a) A physical interaction $\mathcal{H}_{\mathbf{O-S}}$ between physical degrees of freedom regarded as}}
\put(0.5,2){\textit{composing an ``observer'' $\mathbf{O}$ and other, distinct physical degrees of freedom regarded as}}
\put(0.5,1.5){\textit{composing a ``system'' $\mathbf{S}$, all of which are embedded in and interact with physical degrees of}}
\put(0.5,1){\textit{freedom regarded as composing the ``environment'' $\mathbf{E}$.  Boundaries are drawn with broken lines}}
\put(0.5,0.5){\textit{to indicate that they may not be fully characterized by experiments.  (b) A two-way information}}
\put(0.5,0){\textit{transfer between an observer $\mathbf{O}$ and a system $\mathbf{S}$ via a channel $\mathbf{C}$.}}
\end{pspicture}

\psset{xunit=1cm,yunit=1cm}
\begin{pspicture}(0,0)(16,.5)

\end{pspicture}

Replacing ``physical interaction'' with ``informative measurement'' and hence $\mathcal{H}_{\mathbf{O-S}}$ with $\lbrace \mathcal{E}_{\mathit{j}} \rbrace$ effectively replaces Fig. 1a with Fig. 1b, in which a well-defined observer obtains information from a well-defined system.  The surrounding physical environment of Fig. 1a is abstracted into the information channel of Fig. 1b.  This idea that information is transferred from system to observer via the environment is made explicit in quantum Darwinism \cite{zurek06, zurek07grand, zurek09rev}.  However, it is implicit in the assumption of standard decoherence theory that the observer ``ignores'' the surrounding environment and obtains information only from the system; an observer will receive information from the system alone only if the observer - environment interaction transfers no information, i.e. only if the information content of the environment is viewed as transferred entirely through the system - observer channel.  

In the case of human observers of macroscopic systems, the information channel is in many cases physically implemented by the ambient photon field.  If the system of interest is stipulated to be microscopic - the electrons traversing a double-slit apparatus, for example, or a pair of photons in an anti-symmetric Bell state - the information channel is often taken to be the macroscopic measurement apparatus that is employed to conduct the observations.  For the present purposes, the system will be assumed to be macroscopic, and to comprise both the apparatus employed and any additional microscopic degrees of freedom that may be under investigation.  As Fuchs has emphasized \cite{fuchs02, fuchs10}, some intervention in the time-evolution of the system is always required to extract information; hence the channel is two-way as depicted in Fig. 1b.  The fact that the channel delivers \textit{classical} information - real values of pointer variables computed by the component operators of POVMs - imposes on the observer an implicit requirement of classical states into which these classical values may be recorded.  Viewing observation as POVM-mediated information transfer thus requires observers also to be effectively macroscopic.  Consistent with the above characterization of both system and observer as embedded in a ``much larger'' physical environment, the number of states available to either system or observer will be assumed to be much smaller than the number of states within $\mathbf{H}_{\mathbf{U}}$.

Considering the channel through which information flows to be a physical and hence quantum system forcefully raises the question of how the observer identifies as ``$\mathbf{S}$'' the source of the signals that are received.  This is the question that was addressed by Moore \cite{moore56} in the general case of interacting automata.  Moore's answer, that no finite sequence of observations is sufficient to uniquely identify even a classical finite-state machine, calls into question the standard assumption that the observed system can be identified, either by the observer or by a third party, as a collection of physical degrees of freedom represented by a specified set $\lbrace |s_{i}\rangle \rbrace$ of basis vectors.  \textit{Stipulating} that the system can be so represented does not resolve the issue; it merely reformulates the question from one of identifying the system being observed to one of identifying and employing a POVM that acts on the stipulated system and not on something else.  This latter question is eminently practical: it must be addressed in the design of every apparatus and every experimental arrangement. 

By allowing both the degrees of freedom composing the system of interest and the operators composing the POVM employed to perform observations to be arbitrarily stipulated, the standard quantum-mechanical formalism systematically obscures the question of system identification by observers.  While it facilitates computations, placing the ``Heisenberg cut'' delimiting the domain that is to be treated by quantum-mechanical methods around a microscopic collection degrees of freedom further obscures the issue, as it introduces an intermediary - the apparatus - that must also be identified.  It has been shown, moreover, that decoherence considerations alone cannot resolve the question of system identification, as decoherence calculations require the assumption of a boundary that must itself be identified: a boundary in Hilbert space that specifies a collection of degrees of freedom, or a boundary in the space of all possible frameworks or Everett branches that distinguishes the framework or branch under consideration from all others \cite{fields10, fields11a}.  Absent a metaphysical assumption not just of Zurek's axiom(o), but of the specific \textit{a priori} existence of all and only the systems that observers actually observe, the only available sources of such boundary specifications are observers themselves.  The next section examines the question of what such specifications look like in practice.


\section{Informational Requirements for System \\ Identification}

A primary distinction between quantum mechanics and classical mechanics is the failure, in the former but not the latter, of commutativity between physical observables.  Implicit in this statement is the phrase, ``for any given system.''  For example, $[\hat{x}, \hat{p}~] = (\hat{x} \hat{p} - \hat{p} \hat{x}) \neq 0$ says that the position and momentum observables $\hat{x}$ and $\hat{p}$ do not commute for states of any particular, identified system $\mathbf{S}$.  An observation that $\hat{x}$ and $\hat{p}$ do not commute for states of two spatially separated and apparently distinct systems $\mathbf{S}^{\mathrm{1}}$ and $\mathbf{S}^{\mathrm{2}}$ is \textit{prima facie} evidence that $\mathbf{S}^{\mathrm{1}}$ and $\mathbf{S}^{\mathrm{2}}$ are not distinct systems after all.  If $\mathbf{S}^{\mathrm{1}}$ and $\mathbf{S}^{\mathrm{2}}$ are truly distinct, commutativity is not a problem: $[\hat{x}^{\mathrm{1}}, \hat{p}^{\mathrm{2}}] = [\hat{x}^{\mathrm{2}}, \hat{p}^{\mathrm{1}}] = 0$  for all states $|\mathbf{S}^{\mathrm{1}}\rangle$ and $|\mathbf{S}^{\mathrm{2}}\rangle$ operationally defines separability of $\mathbf{S}^{\mathrm{1}}$ from $\mathbf{S}^{\mathrm{2}}$, and warrants the formal representation $|\mathbf{S}^{\mathrm{1}} \otimes \mathbf{S}^{\mathrm{2}}\rangle = |\mathbf{S}^{\mathrm{1}}\rangle \otimes |\mathbf{S}^{\mathrm{1}}\rangle$ of the state of the combined system as separable.  Hence quantum mechanics can only be distinguished from classical mechanics by observers that know when they are observing the same system $\mathbf{S}$ twice, as opposed to observing distinct systems $\mathbf{S}^{\mathrm{1}}$ and $\mathbf{S}^{\mathrm{2}}$, when they test operators for commutativity.

The assumption that a single system $\mathbf{S}$ is being observed is indicated in the standard quantum-mechanical formalism by simply writing down ``$\mathbf{S}$'' and saying: ``Let $\mathbf{S}$ be a physical system ...''   In foundational discussions, however, such a facile and implicit indication of sameness can introduce deep circularity.  Ollivier, Poulin and Zurek, for example, define ``objectivity'' as follows:

\begin{quotation}
``A property of a physical system is \textit{objective} when it is:
\begin{list}{\leftmargin=2em}
\item
1. simultaneously accessible to many observers,
\item
2. who are able to find out what it is without prior knowledge about the system of interest, and 
\item
3. who can arrive at a consensus about it without prior agreement.''
\end{list}
\begin{flushright}
(p. 1 of \cite{zurek04}; p. 3 of \cite{zurek05})
\end{flushright}
\end{quotation}
On the very reasonable assumption that knowing how to identify the system of interest counts as having knowledge about it - exactly what kind of knowledge is discussed in detail below - this definition is clearly circular: each observer must have ``prior knowledge'' to even begin her observations, and the observers must have a ``prior agreement'' that they are observing the same thing to arrive at a consensus about its properties \cite{fields10, fields11a}.  Hence while the assumption that observers \textit{can} know that they are observing one single system over time is natural and even essential to experimentation and practical calculations, both its role as a foundational assumption and its relationship to other assumptions that are explicitly written down as axioms of quantum theory bear examination.

Let us fully specify, therefore, the information that an observer $\mathbf{O}$ must have in order to confirm that $[\mathcal{A}_{\mathrm{1}},\mathcal{A}_{\mathrm{2}}] \neq \mathrm{0}$ for two observables $\mathcal{A}_{\mathrm{1}}$ and $\mathcal{A}_{\mathrm{2}}$ and some physical system $\mathbf{S}$.  The situation can be represented as in Fig. 2: $\mathbf{O}$ is faced with a macroscopic system $\mathbf{S}$, and at any given time $t$ can measure a value for either $\mathcal{A}_{\mathrm{1}}$ or $\mathcal{A}_{\mathrm{2}}$ but not both.  For example, $\mathbf{S}$ could be a Stern-Gerlach apparatus, including ion source, vacuum pump, magnet and power supply, and particle detectors.  In this case, $\mathcal{A}_{\mathrm{1}}$ and $\mathcal{A}_{\mathrm{2}}$ are the spin directions $\hat{s}_{x}$ and $\hat{s}_{z}$, the meters are event counters, and the selector switch sets the position of a mask at either of two fixed angles.  Let us explicitly assume that $\mathbf{O}$ is herself a finite physical system, that $\mathbf{O}$ can make any finite number of measurements in any order, and that $\mathbf{O}$ has been tasked with recording the values for $\mathcal{A}_{\mathrm{1}}$ or $\mathcal{A}_{\mathrm{2}}$ along with the time $t_{\mathit{k}}$ of each observation.  Let us, moreover, explicitly assume that information is physical: that obtaining it requires finite time and recording it requires finite physical memory.  For simplicity, assume also that the information channel $\mathbf{C}$ from $\mathbf{S}$ to $\mathbf{O}$ has sufficient capacity to be regarded as effectively infinite; as this channel is implemented by the environment surrounding the experimental set-up, this assumption is realistic.  

\psset{xunit=1cm,yunit=1cm}
\begin{pspicture}(0,0)(16,6.5)
\pspolygon(5,2)(5,5.5)(11,5.5)(11,2)
\pscircle(6,4){.8}
\psdot(6,4)
\put(6,4){\vector(0,1){.5}}
\put(5.8,2.5){$\mathcal{A}_{\mathrm{1}}$}
\qdisk(8,3){.2}
\pspolygon(8,3.2)(8.4,3.3)(8.2,3)
\put(7.3,3.5){$\mathcal{A}_{\mathrm{1}}$}
\put(8.4,3.5){$\mathcal{A}_{\mathrm{2}}$}
\pscircle(10,4){.8}
\psdot(10,4)
\put(10,4){\vector(1,1){.4}}
\put(9.8,2.5){$\mathcal{A}_{\mathrm{2}}$}

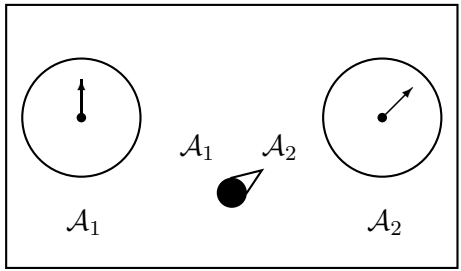
\put(2,0.5){\textit{Fig. 2: A macroscopic system $\mathbf{S}$ with the observable $\mathcal{A}_{\mathrm{2}}$ selected for measurement.}}
\end{pspicture}

Common sense as well as Moore's theorem entail that in order to carry out observations of $\mathbf{S}$, $\mathbf{O}$ must encode information sufficient to (1) distinguish signals from $\mathbf{S}$ from other signals that may flow from the channel; (2) distinguish signals from $\mathbf{S}$ that encode information about the positions of the $\mathcal{A}_{\mathrm{1}} - \mathcal{A}_{\mathrm{2}}$ selector switch and the pointers $\mathbf{P}_{\mathrm{1}}$ and $\mathbf{P}_{\mathrm{2}}$ from signals from $\mathbf{S}$ that do not encode this kind of information; and (3) distinguish between signals that encode different positions of the selector switch and different pointer values for $\mathbf{P}_{\mathrm{1}}$ and $\mathbf{P}_{\mathrm{2}}$.  For example, if $\mathbf{S}$ is a Stern-Gerlach apparatus, $\mathbf{O}$ must encode information sufficient to distinguish $\mathbf{S}$ from other systems of similar size, shape and composition, such as leak detectors or general-purpose mass spectrometers.  Once $\mathbf{O}$ has identified $\mathbf{S}$, she must be capable of identifying the mask selector and the event counters, and determining both the position of the mask and the numbers displayed on the counters.  As $\mathbf{O}$ is finite, all of the information that $\mathbf{O}$ can obtain about $\mathbf{S}$, the selector switch, the pointers, and the values that the pointers indicate can be considered, without loss of generality, to be encoded by finite-precision representations of real numbers.  Assuming that one can talk about a well-defined physical state $|\mathbf{C}\rangle$ of the channel $\mathbf{C}$, the information that $\mathbf{O}$ must encode in order to identify and characterize $\mathbf{S}$ and its components can, therefore, be taken to be encoded by four operators that assign (indicated by ``$\mapsto$'') fine-precision real numbers to states $|\mathbf{C}\rangle$ of $\mathbf{C}$:

\begin{equation*}
\mathcal{S}^{\mathbf{O}}(|\mathbf{C}\rangle) \mapsto \left\{
   \begin{array}{rl}
   (s_{\mathrm{1}}, ..., s_{\mathit{k}}) & \text{if } |\mathbf{C}\rangle \text{ encodes } |\mathbf{S}\rangle \\
   \text{NULL} \quad & \text{otherwise}
   \end{array} \right.
\end{equation*}
where the $s_{\mathrm{1}}, ..., s_{\mathit{k}}$ are finite real values of a set of \textit{control variables} of $\mathbf{S}$;

\begin{equation*}
\mathcal{P}^{\mathbf{O}}(|\mathbf{C}\rangle) \mapsto \left\{
   \begin{array}{rl}
   (p_{\mathrm{1}}, p_{\mathrm{2}}) & \text{if } |\mathbf{C}\rangle \text{ encodes } |\mathbf{S}\rangle \\
   \text{NULL} & \text{otherwise}
   \end{array} \right.
\end{equation*}
where $(p_{\mathrm{1}}, p_{\mathrm{2}}) = (1, 0)$ if the selector switch points to ``$\mathcal{A}_{\mathrm{1}}$'' and $(p_{\mathrm{1}}, p_{\mathrm{2}}) = (0, 1)$ if the selector switch points to ``$\mathcal{A}_{\mathrm{2}}$'';

\begin{equation*}
\mathcal{A}^{\mathbf{O}}_{\mathrm{1}}(|\mathbf{C}\rangle) \mapsto \left\{
   \begin{array}{rl}
   (a_{\mathrm{1 1}} ... a_{\mathrm{1}\mathit{n}}) & \text{if } |\mathbf{C}\rangle \text{ encodes } |\mathbf{P}_{\mathrm{1}}\rangle \\
   \quad & \quad \text{AND } p_{\mathrm{1}} = 1 \\
   \text{NULL} & \text{otherwise}
   \end{array} \right.
\end{equation*}
where $a_{\mathrm{1 1}} ... a_{\mathrm{1}\mathit{n}}$ are finite real values, and;  

\begin{equation*}
\mathcal{A}^{\mathbf{O}}_{\mathrm{2}}(|\mathbf{C}\rangle) \mapsto \left\{
   \begin{array}{rl}
   (a_{\mathrm{2 1}} ... a_{\mathrm{2}\mathit{m}}) & \text{if } |\mathbf{C}\rangle \text{ encodes } |\mathbf{P}_{\mathrm{2}}\rangle \\
   \quad & \quad \text{AND } p_{\mathrm{2}} = 1 \\
   \text{NULL} & \text{otherwise}
   \end{array} \right.
\end{equation*}
where $a_{\mathrm{2 1}} ... a_{\mathrm{2}\mathit{m}}$ are finite real values.  In these expressions, ``NULL'' indicates that the relevant operator returns no value under the indicated conditions.  The allowed values of $a_{\mathrm{1}\mathit{k}}$ and $a_{\mathrm{2}\mathit{k}}$ are the $\mathbf{O}$-distinguishable ``pointer values'' for $\mathcal{A}_{\mathrm{1}}$ and $\mathcal{A}_{\mathrm{2}}$ respectively; they are guaranteed to be both individually finite and finite in number, irrespective of the size of the physical state space of $\mathbf{S}$, by the requirement that a finite observer $\mathbf{O}$ records them with finite precision in a finite memory.  Figure 3 illustrates the action of these operators on $|\mathbf{C}\rangle$, assuming that $\mathbf{S}$ is in the state shown in Fig. 2.

\psset{xunit=1cm,yunit=1cm}
\begin{pspicture}(0,0)(16,7)
\put(.5,6){(a)}
\pspolygon(1.5,2)(1.5,5.5)(7.5,5.5)(7.5,2)
\pscircle(2.5,4){.8}
\psdot(2.5,4)
\put(2.3,2.5){$\mathcal{A}_{\mathrm{1}}$}
\qdisk(4.5,3){.2}
\put(3.8,3.5){$\mathcal{A}_{\mathrm{1}}$}
\put(4.9,3.5){$\mathcal{A}_{\mathrm{2}}$}
\pscircle(6.5,4){.8}
\psdot(6.5,4)
\put(6.3,2.5){$\mathcal{A}_{\mathrm{2}}$}

\put(9,6){(b)}
\pspolygon(10,3.2)(10.4,3.3)(10.2,3)

\put(12,6){(c)}
\put(13,4){\vector(0,1){.5}}
\put(14.5,6){(d)}
\put(15.5,4){\vector(1,1){.4}}

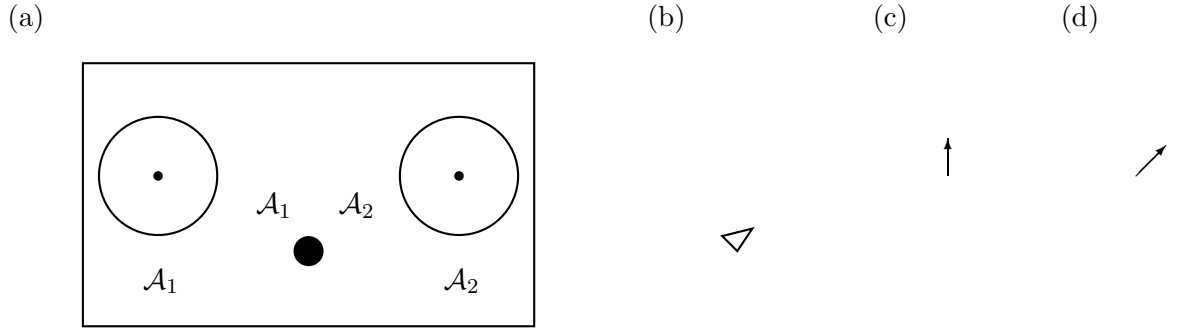
\put(0.5,0.5){\textit{Fig. 3: State information assigned by the operators (a) $\mathcal{S}^{\mathbf{O}}$, (b) $\mathcal{P}^{\mathbf{O}}$, (c) $\mathcal{A}^{\mathbf{O}}_{\mathrm{1}}$, and (d) $\mathcal{A}^{\mathbf{O}}_{\mathrm{2}}$ on $|\mathbf{C}\rangle$.}}
\put(0.5,0){\textit{The operator $\mathcal{S}^{\mathbf{O}}$ assigns state information about all components of $\mathbf{S}$ other than the selector}}
\end{pspicture}

\psset{xunit=1cm,yunit=1cm}
\begin{pspicture}(0,0)(16,2)
\put(0.5,1.5){\textit{switch and pointers.  The operator $\mathcal{P}^{\mathbf{O}}$ assigns state information about the selector switch only.}}
\put(0.5,1){\textit{The operators $\mathcal{A}^{\mathbf{O}}_{\mathrm{1}}$ and  $\mathcal{A}^{\mathbf{O}}_{\mathrm{2}}$, respectively, assign state information about the positions of the left-}} \put(0.5,0.5){\textit{and right-hand pointers only.}}
\end{pspicture}

As illustrated in Fig. 3, the values of the control variables $s_{\mathrm{1}}, ..., s_{\mathit{k}}$ are what indicate to $\mathbf{O}$ that she is in fact observing $\mathbf{S}$ and not something else.  In the case of the Stern-Gerlach apparatus, these may include details of its size, shape and components, as well as conventional symbols such as brand names or read-out labels.  In order for $\mathbf{O}$ to recognize these values, they clearly must be real and finite. The control variables must, moreover, take on ``acceptable'' values at $t$ indicating to $\mathbf{O}$ that $\mathbf{S}$ is in a state suitable for making observations.  A Stern-Gerlach apparatus, for example, must have an acceptable value for the chamber vacuum and the magnets and particle detectors must be turned on.  The entire apparatus must not be disassembled, under repair, or on fire.  The existence, recognition by the observer, and acceptable values of such control variables are being assumed whenever ``$\mathbf{S}$'' is written down as the name of a quantum system that is being observed.  It is commonplace in the literature (e.g. \cite{tegmark00} where this is explicit) to treat quantum systems as represented during the measurement process by their pointer states alone, but as Figs. 3c and 3d illustrate, such a ``bare pointer'' provides no information by which the system for which it indicates a pointer value can be identified, much less be determined to be in an acceptable state for making observations. 

The operators $\mathcal{S}^{\mathbf{O}}$, $\mathcal{P}^{\mathbf{O}}$, $\mathcal{A}^{\mathbf{O}}_{\mathrm{1}}$ and $\mathcal{A}^{\mathbf{O}}_{\mathrm{2}}$ defined above assign finite values, i.e. do not assign ``NULL,'' only for subsets of the complete set of states of $\mathbf{C}$.  As discussed above, the information channel $\mathbf{C}$ is physically implemented by the environment in which $\mathbf{S}$ and $\mathbf{O}$ are embedded.  Let $\mathbf{H}_{\mathbf{C}}$ be the Hilbert space of this environment.  As the environment of any experiment is contiguous with the universe as a whole, with increasing elapsed time the dimension $dim(\mathbf{H}_{\mathbf{C}}) \sim \mathit{dim}(\mathbf{H}_{\mathbf{U}})$; $\mathbf{H}_{\mathbf{C}}$ can therefore be considered to be much larger than the state spaces of either $\mathbf{S}$ or $\mathbf{O}$, and in particular much larger than the memory available to $\mathbf{O}$.  Let $\mathcal{S}^{\mathbf{O}}_{\mathit{NULL}}$, $\mathcal{P}^{\mathbf{O}}_{\mathit{NULL}}$, $\mathcal{A}^{\mathbf{O}}_{\mathrm{1} \mathit{NULL}}$ and $\mathcal{A}^{\mathbf{O}}_{\mathrm{2} \mathit{NULL}}$, respectively, be operators defined on $\mathbf{H}_{\mathbf{C}}$ that assign a value of zero to all states within $\mathbf{H}_{\mathbf{C}}$ that do not encode information about the states of $\mathbf{S}$, the selector switch of $\mathbf{S}$, $\mathbf{P}_{\mathrm{1}}$ and $\mathbf{P}_{\mathrm{2}}$ respectively, and ``NULL'' for states within $\mathbf{H}_{\mathbf{C}}$ that do encode such information.  A POVM $\lbrace \mathcal{S}^{\mathbf{O}}_{\mathit{k}} \rbrace$ acting on $\mathbf{H}_{\mathbf{C}}$ can then be defined as follows: let $\mathcal{S}^{\mathbf{O}}_{\mathit{0}} = \mathcal{S}^{\mathbf{O}}_{\mathit{NULL}}$, and for $k \neq 0$ let $\mathcal{S}^{\mathbf{O}}_{\mathit{k}}$ be the component of $\mathcal{S}^{\mathbf{O}}$ that assigns the value $s_{k}$, normalized so that $\mathcal{S}^{\mathbf{O}}_{\mathit{0}} + \sum_{\mathit{k \neq 0}} \mathcal{S}^{\mathbf{O}}_{\mathit{k}} = \mathit{Id}$ where $Id$ is the identity operator for $\mathbf{H}_{\mathbf{C}}$.  The component $\mathcal{S}^{\mathbf{O}}_{\mathit{0}}$ of $\lbrace \mathcal{S}^{\mathbf{O}}_{\mathit{k}} \rbrace$ is by definition orthogonal to the $\mathcal{S}^{\mathbf{O}}_{\mathit{k}}$ with $k \neq 0$; however, these latter components are not, in general, required to be orthogonal to each other.  The component of $\lbrace \mathcal{S}^{\mathbf{O}}_{\mathit{k}} \rbrace$ that assigns the value ``ready'' to $\mathbf{S}$, for example, will not in general be orthogonal to components that establish the identity of $\mathbf{S}$; many parts of $\mathbf{S}$ must be examined to determine that it is ready for use.  Practical experimental apparatus are, nonetheless, generally designed to assure that many non-NULL components of $\lbrace \mathcal{S}^{\mathbf{O}}_{\mathit{k}} \rbrace$ are orthogonal and hence distinguishable and informationally independent.  The vacuum gauge on a Stern-Gerlach apparatus, for example, is designed to be distinguishable from and independent of the ammeter on the magnet power supply or the readout on the event counter.  In general, the distinguishability and informational independence of components is an operational definition of their separability and hence of the appearance of classicality.  The practical requirement that observer-identifiable systems have distinguishable and informationally-independent control and pointer variables is analogous to Bohr's requirement \cite{bohr28} that measurement apparatus be regarded as classical.

Additional POVMs $\lbrace \mathcal{P}^{\mathbf{O}}_{\mathit{k}} \rbrace$, $\lbrace \mathcal{A}^{\mathbf{O}}_{\mathrm{1}\mathit{k}} \rbrace$ and $\lbrace \mathcal{A}^{\mathbf{O}}_{\mathrm{2}\mathit{k}} \rbrace$ can be defined by including $\mathcal{P}^{\mathbf{O}}_{\mathit{NULL}}$, $\mathcal{A}^{\mathbf{O}}_{\mathrm{1} \mathit{NULL}}$ and $\mathcal{A}^{\mathbf{O}}_{\mathrm{2} \mathit{NULL}}$ as $0^{th}$ components.  As in the case of $\lbrace \mathcal{S}^{\mathbf{O}}_{\mathit{k}} \rbrace$, these $0^{th}$ components are by definition orthogonal to the others.  If $\mathbf{S}$ is assumed to be designed so as to allow only a single kind of measurement to be performed at any given time, and if all observations are assumed to be carried out at maximum resolution, then the non-NULL components of $\lbrace \mathcal{P}^{\mathbf{O}}_{\mathit{k}} \rbrace$, $\lbrace \mathcal{A}^{\mathbf{O}}_{\mathrm{1}\mathit{k}} \rbrace$ and $\lbrace \mathcal{A}^{\mathbf{O}}_{\mathrm{2}\mathit{k}} \rbrace$ can also be taken to be orthogonal.  For simplicity, orthogonality of these components will be assumed in what follows; the general case can be accomodated by assuming that the components of $\lbrace \mathcal{P}^{\mathbf{O}}_{\mathit{k}} \rbrace$ that indicate incompatible measurements are orthogonal, that components of $\lbrace \mathcal{A}^{\mathbf{O}}_{\mathrm{1}\mathit{k}} \rbrace$ and $\lbrace \mathcal{A}^{\mathbf{O}}_{\mathrm{2}\mathit{k}} \rbrace$ that assign values at maximum resolution are orthogonal, and by considering only these orthogonal components when defining inverse images as described below.  

Regarding $\mathcal{S}^{\mathbf{O}}$, $\mathcal{P}^{\mathbf{O}}$, $\mathcal{A}^{\mathbf{O}}_{\mathrm{1}}$ and $\mathcal{A}^{\mathbf{O}}_{\mathrm{2}}$ respectively as POVMs $\lbrace \mathcal{S}^{\mathbf{O}}_{\mathit{k}} \rbrace$, $\lbrace \mathcal{P}^{\mathbf{O}}_{\mathit{k}} \rbrace$, $\lbrace \mathcal{A}^{\mathbf{O}}_{\mathrm{1}\mathit{k}} \rbrace$ and $\lbrace \mathcal{A}^{\mathbf{O}}_{\mathrm{2}\mathit{k}} \rbrace$ acting on $\mathbf{H}_{\mathbf{C}}$ is useful because it removes any dependence on an explicit specification of the boundaries between $\mathbf{C}$ and $\mathbf{S}$ or between the selector switch, $\mathbf{P}_{\mathrm{1}}$, $\mathbf{P}_{\mathrm{2}}$ and the non-switch and non-pointer components of $\mathbf{S}$.  These boundaries are replaced, from $\mathbf{O}$'s perspective, by the boundaries of the $\mathbf{O}$-detectable encodings of $\mathbf{S}$ and its components in $\mathbf{H}_{\mathbf{C}}$.  Let $\epsilon$ be $\mathbf{O}$'s detection threshold for encodings in $\mathbf{C}$; $\mathbf{O}$ is able to record a value $s_{k}$, for example, only if $\langle \mathbf{C}|\mathcal{S}^{\mathbf{O}}_{\mathit{k}}| \mathbf{C}\rangle \geq \epsilon$.  Because $\mathbf{O}$ is a finite observer, $\epsilon > 0$; arbitrarily weak encodings are not detectable.  Given this threshold, the encoding of $\mathbf{S}$ can be defined, from $\mathbf{O}$'s perspective, as $\cup_{k} (Im^{-1}(s_{k}))$, where $Im^{-1}(s_{k})$ is the inverse image in $\mathbf{H}_{\mathbf{C}}$ of the detectable value $s_{k}$.  Because $\mathcal{S}^{\mathbf{O}}_{\mathit{0}}$ is orthogonal to all of the $\mathcal{S}^{\mathbf{O}}_{\mathit{k}}$ with $k \neq 0$, the intersection $Im^{-1} (\mathcal{S}^{\mathbf{O}}_{\mathit{0}}) \mathit{\cap (\cup_{k} (Im^{-1}(s_{k}))) = \emptyset}$; indeed these inverse images are separated by states for which $0 \leq \langle \mathbf{C}|\mathcal{S}^{\mathbf{O}}_{\mathit{k}}| \mathbf{C}\rangle \leq \epsilon$ for all $\mathcal{S}^{\mathbf{O}}_{\mathit{k}}$ with $k \neq 0$.  Let ``$Im^{-1} \lbrace \mathcal{S}^{\mathbf{O}}_{\mathit{k}} \rbrace$'' denote $\cup_{k} (Im^{-1}(s_{k}))$; $Im^{-1} \lbrace \mathcal{S}^{\mathbf{O}}_{\mathit{k}} \rbrace$ is then the proper subspace of $\mathbf{H}_{\mathbf{C}}$ containing vectors to which the POVM $\lbrace \mathcal{S}^{\mathbf{O}}_{\mathit{k}} \rbrace$ assigns finite real values with probabilities greater than $\epsilon$.  The proper subspaces $Im^{-1} \lbrace \mathcal{P}^{\mathbf{O}}_{\mathit{k}} \rbrace$, $Im^{-1} \lbrace \mathcal{A}^{\mathbf{O}}_{\mathrm{1}\mathit{k}} \rbrace$ and $Im^{-1} \lbrace \mathcal{A}^{\mathbf{O}}_{\mathrm{1}\mathit{k}} \rbrace$ can be defined in an analogous fashion.  As any state $|C\rangle$ that encodes an acceptable value of either the pointer position or the pointer value for either $\mathcal{A}^{\mathbf{O}}_{\mathrm{1}}$ or $\mathcal{A}^{\mathbf{O}}_{\mathrm{2}}$ also encodes acceptable values of the $s_{k}$, it is clear that $Im^{-1} \lbrace \mathcal{P}^{\mathbf{O}}_{\mathit{k}} \rbrace$, $Im^{-1} \lbrace \mathcal{A}^{\mathbf{O}}_{\mathrm{1}\mathit{k}} \rbrace$ and $Im^{-1} \lbrace \mathcal{A}^{\mathbf{O}}_{\mathrm{1}\mathit{k}} \rbrace$ are properly contained within $Im^{-1} \lbrace \mathcal{S}^{\mathbf{O}}_{\mathit{k}} \rbrace$.

Specifying $\lbrace \mathcal{S}^{\mathbf{O}}_{\mathit{k}} \rbrace$, $\lbrace \mathcal{P}^{\mathbf{O}}_{\mathit{k}} \rbrace$, $\lbrace \mathcal{A}^{\mathbf{O}}_{\mathrm{1}\mathit{k}} \rbrace$ and $\lbrace \mathcal{A}^{\mathbf{O}}_{\mathrm{2}\mathit{k}} \rbrace$ in terms of the values that they assign for each state $|\mathbf{C}\rangle$ of $\mathbf{C}$ completely specifies $\mathbf{O}$'s observational capabilities regarding $\mathbf{S}$; no further specification of $\mathbf{S}$ or its states is necessary.  The notion that ``systems exist'' can, therefore, be dropped; all that is necessary for the description of measurement, other than observers equipped with POVMs, is that \textit{channels} exist.  By regarding all POVMs that identify systems or their components as observer-specific (hence dropping the superscript ``$\mathbf{O}$''), the minimal capabilities required by any observer can be defined in purely information-theoretic terms.  Given an information channel $\mathbf{C}$, a \textit{minimal observer} on $\mathbf{C}$ is a finite system $\mathbf{O}$ that encodes collections of POVMs $\lbrace \mathcal{S}^{\mathit{i}}_{\mathit{k}} \rbrace$, $\lbrace \mathcal{P}^{\mathit{i}}_{\mathit{k}} \rbrace$ and $\lbrace \mathcal{A}^{\mathit{i}}_{\mathit{jk}} \rbrace$ within a control structure such that, for each $i$:

\begin{enumerate}
\item The inverse images $Im^{-1} \lbrace \mathcal{S}^{\mathit{i}}_{\mathit{k}} \rbrace$, $Im^{-1} \lbrace \mathcal{P}^{\mathit{i}}_{\mathit{k}} \rbrace$ and $Im^{-1} \lbrace \mathcal{A}^{\mathit{i}}_{\mathit{jk}} \rbrace$ for all $j$ are non-empty proper subspaces of $\mathbf{H}_{\mathbf{C}}$ such that $Im^{-1} \lbrace \mathcal{S}^{\mathit{i}}_{\mathit{k}} \rbrace$ properly contains $Im^{-1} \lbrace \mathcal{P}^{\mathit{i}}_{\mathit{k}} \rbrace$ and the $Im^{-1} \lbrace \mathcal{A}^{\mathit{i}}_{\mathit{jk}} \rbrace$ for all $j$.
\item  The $s^{\mathit{i}}_{\mathrm{1}}, ..., s^{\mathit{i}}_{n^{\mathit{i}}}$ are accepted by the control structure of $\mathbf{O}$ as triggering the action of the POVM $\lbrace \mathcal{A}^{\mathit{i}}_{\mathit{jk}} \rbrace$ for which $p^{\mathit{i}}_{j} = 1$.
\item The control structure of $\mathbf{O}$ is such that the action on $|\mathbf{C}\rangle$ with $\mathcal{A}^{\mathit{i}}_{\mathit{jk}}$ is followed by recording of the single non-zero value $a^{\mathit{i}}_{\mathit{jk}}$ to memory. 
\end{enumerate}
The control structure required by this definition consists of one ``if - then - else'' block for each POVM component, organized as shown in Fig. 4 for a minimal observer with $N$ POVMs $\lbrace \mathcal{S}^{\mathit{i}}_{\mathit{k}} \rbrace$.  Together with the specified POVMs and a memory allocation process, this control structure specifies a classical virtual machine (e.g. \cite{tan76}), i.e. a consistent semantic interpretation of some subset of the possible behaviors of a computing device.  Such a virtual machine may be implemented as software on any Turing-equivalent functional architecture, and hence may be physically implemented by any quantum system that provides a Turing-equivalent functional architecture, such as a QTM \cite{deutsch85} or any of the alternative quantum computing architectures provably equivalent to a QTM \cite{nielsen-chaung00, farhi96, galindo01, perdrix-jorrand06}.  \textit{Constructing} such an implementation using a programming language provided by a quantum computing architecture is equivalent to constructing a semantic interpretation of the behavior of the quantum computing architecture that defines the virtual machine using the pre-defined semantics of the programming language.  As in the classical case, programming languages for quantum computing architectures provide the required semantic mappings from formal computational constructs (e.g. logical operations or arithmetic) to the operations of the underlying architecture (e.g. unitary dynamics for a QTM or a Hamiltonian oracle) \cite{gay06, rudiger07}; for any universal programming language, however, higher-level interpretations that define specific programs are independent of these lower-level semantic mappings.  Hence from an ontological perspective, a minimal observer is a \textit{classical virtual machine} that is \textit{physically implemented} by a quantum system $\mathbf{O}$ that, if not universal, nonetheless provides a sufficient quantum computing architecture to realize all the functions of the minimal observer.  A physically-implemented minimal observer interacts with and obtains physically-encoded information from a physically implemented information channel $\mathbf{C}$.  Laboratory data acquisition systems that incorporate signal-source identification criteria and stably record measurement results are minimal observers under this definition.  As is the case for all physical implementations of classical virtual machines, and for all operations involving classically-characterized inputs to or outputs from quantum computers, the semantic interpretation of a physical (i.e. quantum) system as an implementation of a minimal observer requires, at least implicitly via the semantics of the relevant programming language, an interpretative approach to the quantum measurement problem.  The consequences of replacing Galilean observers with minimal observers as defined here for interpretative approaches to the measurement problem are discussed in Sect. 6 below.

\psset{xunit=1cm,yunit=1cm}
\begin{pspicture}(0,0)(16,16)
\put(15.5,15.1){\line(-1,0){12}}

\put(3.5,15.1){\vector(0,-1){1.1}}
\pspolygon(2,13)(3.5,14)(5,13)(3.5,12)
\put(2.9,13.2){Accept}
\put(2.6,12.7){$s^{\mathrm{1}}_{\mathrm{1}}, ..., s^{\mathrm{1}}_{n^{\mathrm{1}}}$?}
\put(5,13){\line(1,0){1.5}}
\psdot(6.9,13)
\psdot(7.3,13)
\psdot(7.7,13)
\put(8,13){\line(1,0){1.5}}
\put(5.5,13.2){No}
\put(3.5,12){\vector(0,-1){1}}
\pspolygon(2,10)(3.5,11)(5,10)(3.5,9)
\put(2.8,10){$p^{\mathrm{1}}_{\mathrm{1}} = 1$?}
\put(5,10){\line(1,0){1.5}}
\put(5.5,10.2){No}
\put(6.5,10){\line(0,-1){.3}}
\psdot(6.5,9.3)
\psdot(6.5,9)
\psdot(6.5,8.7)
\put(6.5,8.3){\vector(0,-1){.3}}
\pspolygon(5,7)(6.5,8)(8,7)(6.5,6)
\put(5.8,7){$p^{\mathrm{1}}_{m^{\mathrm{1}}} = 1$?}
\put(8,7){\line(1,0){1.4}}
\put(9.6,7){\line(1,0){2.8}}
\put(12.6,7){\vector(1,0){2.9}}
\put(8.5,7.2){No}
\psdot(15.5,7)
\put(9.5,13){\vector(0,-1){.5}}
\pspolygon(8,11.5)(9.5,12.5)(11,11.5)(9.5,10.5)
\put(8.9,11.8){Accept}
\put(8.6,11.3){$s^{N}_{\mathrm{1}}, ..., s^{N}_{n^{N}}$?}
\put(11.5,11.7){No}
\put(11,11.5){\vector(1,0){4.5}}
\psdot(15.5,11.5)
\put(9.5,10.5){\vector(0,-1){.7}}

\pspolygon(8,8.8)(9.5,9.8)(11,8.8)(9.5,7.8)
\put(8.7,8.8){$p^{N}_{\mathrm{1}} = 1$?}
\put(11,8.8){\line(1,0){1.5}}
\put(11.5,9){No}
\put(12.5,8.8){\line(0,-1){.3}}
\psdot(12.5,8.2)
\psdot(12.5,7.9)
\psdot(12.5,7.6)
\put(12.5,7.3){\vector(0,-1){.5}}
\pspolygon(11,5.8)(12.5,6.8)(14,5.8)(12.5,4.8)
\put(11.7,5.7){$p^{N}_{m^{N}} = 1$?}
\put(14,5.8){\vector(1,0){1.5}}
\put(14.5,6){No}
\psdot(15.5,5.8)
\put(3.5,9){\vector(0,-1){0.5}}
\pspolygon(2,7.5)(2,8.5)(5,8.5)(5,7.5)
\put(2.2,7.9){Record $a^{\mathrm{1}}_{\mathrm{1}\mathit{k}} \neq 0$}
\put(3.5,7.5){\vector(0,-1){5.5}}
\put(6.5,6){\vector(0,-1){.5}}
\pspolygon(4.9,4.5)(4.9,5.5)(8,5.5)(8,4.5)
\put(5,4.9){Record $a^{\mathrm{1}}_{m^{\mathrm{1}}k} \neq 0$}
\put(6.5,4.5){\vector(0,-1){2.5}}
\put(9.5,7.8){\vector(0,-1){3}}
\pspolygon(8.1,3.8)(8.1,4.8)(11.1,4.8)(11.1,3.8)
\put(8.3,4.1){Record $a^{N}_{\mathrm{1}\mathit{k}} \neq 0$}
\put(9.5,3.8){\vector(0,-1){1.8}}
\put(12.5,4.8){\vector(0,-1){1.3}}
\pspolygon(10.9,2.5)(10.9,3.5)(14.1,3.5)(14.1,2.5)
\put(11,2.8){Record $a^{N}_{m^{N}k} \neq 0$}
\put(12.5,2.5){\vector(0,-1){.5}}
\pspolygon(3,1)(3,2)(13,2)(13,1)
\put(5.6,1.4){Allocate new memory block}
\put(13,1.5){\line(1,0){2.5}}
\put(15.5,1.5){\line(0,1){13.6}}

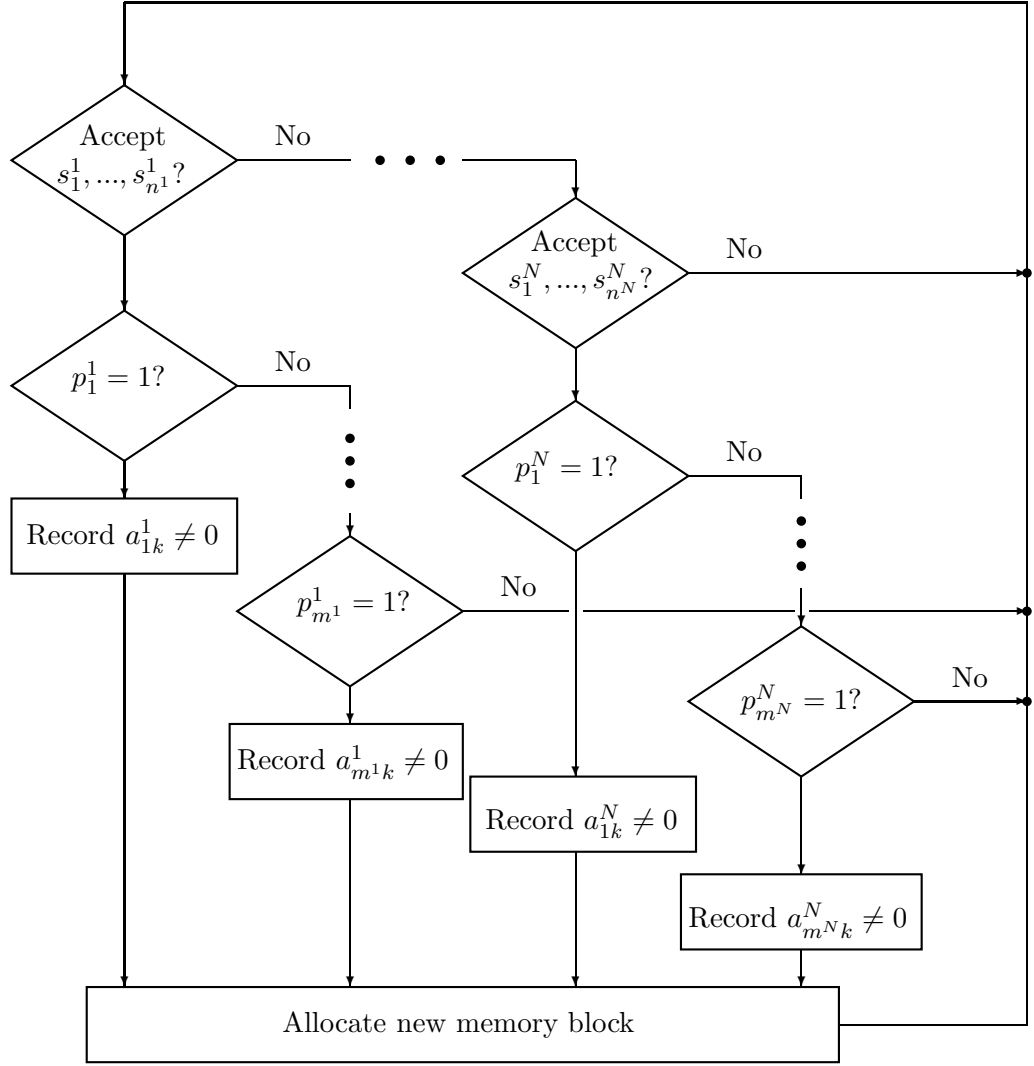
\put(0.5,0.3){\textit{Fig. 4: Organization of ``if - then - else'' blocks in the control structure of a minimal observer.}}
\end{pspicture}

A minimal observer as defined above, and as illustrated in Fig. 4, is clearly not Galilean; it is rather a richly-structured information-encoding entity.  The information encoded by a minimal observer is \textit{relative to} a specified control structure, and is therefore \textit{pragmatic}, i.e. used for doing something \cite{roederer05, roederer11}.  Hence a minimal observer is not just a ``physical system having a definite state of motion'' or ``a quantum system interacting with the observed system.''  Indeed, if considered apart from its physical implementation, a minimal observer as defined above is not a \textit{quantum system} at all; it is a classical virtual machine, an entity defined purely informationally.  One cannot, therefore, talk about the ``quantum state'' of a minimal observer.  The traditional von Neumann chain representation (\cite{vonNeumann32}, reviewed e.g. by \cite{schloss07}), in which the observer becomes entangled with the system of interest, after which the observer's quantum state must ``collapse'' to a definite outcome, cannot be defined for a minimal observer, and the information encoded by a minimal observer cannot be characterized by a von Neumann entropy.  The \textit{physical implementation} of a minimal observer can be characterized by a quantum state, and hence does have a von Neumann entropy; however, \textit{any} physical implementation that provides a Turing-equivalent architecture and sufficient coding capacity will do.  The history of compilers, interpreters, programming languages, and distributed architecures demonstrates that the emulation mapping from a virtual machine to its physical implementation can be arbitrarily complex, indirect, and de-localized in space and time; any straightforward interpretation of von Neumann's principle of ``psychophysical parallelism'' as a constraint on the implementation of minimal observers is, therefore, undone by the architecture that von Neumann himself helped devise two decades after the publication of \textit{Mathematische Grundlagen der Quantenmechanische}.

In consequence of their finite supplies of executable POVMs and finite memories, minimal observers display \textit{objective ignorance} of two distinct kinds.  First, a minimal observer cannot, by any finite sequence of observations, fully specify the set of states of $\mathbf{C}$ that encode states of any system $\mathbf{S}$, regardless of the size of the state space of $\mathbf{S}$.  This form of objective ignorance follows solely from the large size of $\mathbf{H}_{\mathbf{C}}$ compared to memory available to $\mathbf{O}$.  A minimal observer cannot, therefore, determine with certainty that any specification of the states of $\mathbf{S}$ derived from observations is complete.  If the observational data characterizing $\mathbf{S}$ obtained by $\mathbf{O}$ are viewed as outputs from an oracle, this failure of completeness can be viewed as an instance of the Halting Problem \cite{tan76, hopcroft79}: $\mathbf{O}$ cannot, in principle, determine whether any oracle that produces a specification of the states of $\mathbf{S}$ will halt in finite time.  This first form of objective ignorance blocks for minimal observers the standard assumption of particle physics that the states of elementary particles are specified completely by their observable quantum numbers, downgrading this to a ``for all practical purposes'' specification; it then extends this restriction to all systems, elementary or not.  The second form of objective ignorance is that required by Moore's theorem: any system $\mathbf{S}^{\prime}$ that interacts with $\mathbf{C}$ in a way that is indistinguishable using $\lbrace \mathcal{S}^{\mathbf{O}}_{\mathit{k}} \rbrace$, $\lbrace \mathcal{P}^{\mathbf{O}}_{\mathit{k}} \rbrace$, $\lbrace \mathcal{A}^{\mathbf{O}}_{\mathrm{1}\mathit{k}} \rbrace$ and $\lbrace \mathcal{A}^{\mathbf{O}}_{\mathrm{2}\mathit{k}} \rbrace$ from $\mathbf{S}$ will be identified by $\mathbf{O}$ as $\mathbf{S}$.  The information provided to $\mathbf{O}$ by $\lbrace \mathcal{S}^{\mathbf{O}}_{\mathit{k}} \rbrace$, $\lbrace \mathcal{P}^{\mathbf{O}}_{\mathit{k}} \rbrace$, $\lbrace \mathcal{A}^{\mathbf{O}}_{\mathrm{1}\mathit{k}} \rbrace$ and $\lbrace \mathcal{A}^{\mathbf{O}}_{\mathrm{2}\mathit{k}} \rbrace$ is, therefore, objectively ambiguous concerning the physical degrees of freedom that generate the encodings in $\mathbf{C}$ on which these operators act.  This second form of objective ignorance extends to all systems the indistinguishability within types familiar from particle physics.  Neither of these forms of objective ignorance can be remedied by further data acquisition by $\mathbf{O}$; they thus differ fundamentally from subjective or classical ignorance.  As will be shown in the two sections that follow, these two forms of objective ignorance together assure that the observational results recorded by a minimal observer will display the typical characteristics predicted by quantum theory, independently of any specific assumptions about the observer's physical implementation.

\section{Physical Interpretation of Non-commutative POVMs}

The definition of a minimal observer given above relies only on the classical concept of information and the system-identification requirements placed on observers by classical automata theory, the assumption that the channel $\mathbf{C}$ is physically implemented by the environment, the idea that information is physical, and the formal notion of a POVM.  It provides, however, a robust formal framework with which arbitrary measurement interactions can be characterized.  This formal framework makes no mention of ``systems'' other than $\mathbf{O}$ and $\mathbf{C}$, requires no strict specification of the boundary between the physical degrees of freedom that implement $\mathbf{O}$ and those that implement $\mathbf{C}$, and makes no assumption that $\mathbf{O}$ and $\mathbf{C}$ are separable.  The physical interpretation of information transfer by POVMs within this framework thus provides a ``systems-free'' interpretation of quantum mechanics with no \textit{a priori} assumptions about the nature of quantum states.  This interpretation does not violate the axiomatic assumptions of minimal quantum mechanics in any way; hence it requires no changes in the standard quantum-mechanical formalism or its application in practice to specific cases.

Let us drop temporarily the assumption of minimal quantum mechanics adopted in Sect. 2, and assume only that the physical degrees of freedom composing the coupled system $\mathbf{O \otimes C}$, where ``$\mathbf{O}$'' here refers to the physical implementation of a minimal observer, evolve under some dynamics $\mathcal{H}$ that is time-symmetric and fully deterministic.  A natural, classical ``arrow of time'' is imposed on this dynamics, from the perspective of $\mathbf{O}$, by the sequence of memory allocations executed by $\mathbf{O}$'s functional architecture.  From a perspective exterior to $\mathbf{O}$ (e.g. the perspective of $\mathbf{C}$), the minimal observer $\mathbf{O}$ is only one of an arbitrarily large number of virtual machines that could describe the physical dynamics of its hardware implementation; hence this $\mathbf{O}$-specific arrow of time is unavailable from such an exterior perspective.  Any alternative minimal observer $\mathbf{O}^{\prime}$ will, however, have its own arrow of time determined by its own memory-allocation process.

The large size of $\mathbf{H}_{\mathbf{C}}$ renders the physical degrees of freedom implementing $\mathbf{C}$ \textit{fine-grained} compared to both the detection resolution $\epsilon$ and the memory capacity of any minimal observer $\mathbf{O}$; in particular, these degrees of freedom are fine-grained compared to the inverse images of the POVMs $\lbrace \mathcal{S}^{\mathit{i}}_{\mathit{k}} \rbrace$, $\lbrace \mathcal{P}^{\mathit{i}}_{\mathit{k}} \rbrace$ and $\lbrace \mathcal{A}^{\mathit{i}}_{\mathit{jk}} \rbrace$ with which $\mathbf{O}$ obtains information about an external system $\mathbf{S}$.  As illustrated in Fig. 1a, $\mathbf{O}$ is implemented by the same kinds of physical degrees of freedom that implement $\mathbf{C}$; the degrees of freedom implementing $\mathbf{O}$ are, therefore, also fine-grained compared to $\mathbf{O}$'s memory.  Let us assume a weak version of counterfactual definiteness: that the fine-grained degrees of freedom within the inverse images of the $\lbrace \mathcal{S}^{\mathit{i}}_{\mathit{k}} \rbrace$, $\lbrace \mathcal{P}^{\mathit{i}}_{\mathit{k}} \rbrace$ and $\lbrace \mathcal{A}^{\mathit{i}}_{\mathit{jk}} \rbrace$ implemented by any $\mathbf{O}$ are well-defined at all times; this assumption is a natural correlate, if not a consequence, of the realist stance toward physical degrees of freedom adopted in Sect. 2.  Note that this assumption of counterfactual definiteness does not apply to the states of any ``system'' other than $\mathbf{C}$, and that it applies to states of $\mathbf{C}$ without assuming that $\mathbf{C}$ is separable from $\mathbf{O}$.  This assumption renders any physical interpretation based on it a ``hidden variables'' theory.  However, it does not violate the Kochen-Specker contextuality theorem \cite{kochen67}; indeed it provides a mechanism for satisfying it.  The ``hidden'' fine-grained state variables of $\mathbf{C}$ are inaccessible in principle to $\mathbf{O}$, although they fully determine the course-grained measurement results that $\mathbf{O}$ obtains.  As discussed above, no two instances of the execution of a $\lbrace \mathcal{S}^{\mathit{i}}_{\mathit{k}} \rbrace$, $\lbrace \mathcal{P}^{\mathit{i}}_{\mathit{k}} \rbrace$, $\lbrace \mathcal{A}^{\mathit{i}}_{\mathit{jk}} \rbrace$ triple at times $t$ and $t^{\prime}$ can be assumed by $\mathbf{O}$ to act on the same fine-grained state $|\mathbf{C}\rangle$, nor is any measure of similarity or dissimilarity of channel states $|\mathbf{C}\rangle$ and $|\mathbf{C}\rangle^{\prime}$ other than a $\lbrace \mathcal{S}^{\mathit{i}}_{\mathit{k}} \rbrace$, $\lbrace \mathcal{P}^{\mathit{i}}_{\mathit{k}} \rbrace$, $\lbrace \mathcal{A}^{\mathit{i}}_{\mathit{jk}} \rbrace$ triple available to $\mathbf{O}$.  All executions by $\mathbf{O}$ of a single measurement $\lbrace \mathcal{A}^{\mathit{i}}_{\mathit{jk}} \rbrace$ are thus contextualized by prior executions of $\lbrace \mathcal{S}^{\mathit{i}}_{\mathit{k}} \rbrace$ and $\lbrace \mathcal{P}^{\mathit{i}}_{\mathit{k}} \rbrace$; executions of pairs $\lbrace \mathcal{A}^{\mathit{i}}_{\mathit{jk}} \rbrace$ and $\lbrace \mathcal{A}^{\mathit{i}}_{\mathit{lm}} \rbrace$, commutative or otherwise, are contextualized by two executions of $\lbrace \mathcal{S}^{\mathit{i}}_{\mathit{k}} \rbrace$ and $\lbrace \mathcal{P}^{\mathit{i}}_{\mathit{k}} \rbrace$.

Finally, let us assume that the dynamic evolution of $\mathbf{C}$ does not depend in any way on the POVMs or the control structure implemented by $\mathbf{O}$.  Given that $\mathbf{O}$ is by definition a virtual machine, this is an assumption that the physical dynamics $\mathcal{H}$ is independent of its semantic interpretation by any observer.  This assumption of \textit{decompositional equivalence} assures that the allocation by $\mathcal{H}$ of fine-grained degrees of freedom to the inverse images of the $\lbrace \mathcal{S}^{\mathit{i}}_{\mathit{k}} \rbrace$, $\lbrace \mathcal{P}^{\mathit{i}}_{\mathit{k}} \rbrace$ and $\lbrace \mathcal{A}^{\mathit{i}}_{\mathit{jk}} \rbrace$ are independent of the information $\mathbf{O}$ encodes, and hence of $\mathbf{O}$'s ``expectations'' about $\mathbf{C}$ or $\mathcal{H}$.  This assumption renders the interpretative framework free of ``subjective'' dependence on the observer.  By ruling out any dependence of $\mathcal{H}$ on system - environment boundaries drawn by observers, it also renders the interpretative framework consistent with the common scientific practice of stipulating systems of interest \textit{ad hoc} either demonstratively by pointing and saying ``that'' or formally by specifying lists of degrees of freedom to be included within the boundaries of the stipulated systems.

With these assumptions, the interpretation of $\mathbf{O \otimes C}$ is both realist and objectivist about the fine-grained degrees of freedom implementing $\mathbf{O \otimes C}$, and free, via decompositional equivalence, of any dependence on what observables and hence what descriptions of $\mathbf{C}$ or $\mathcal{H}$ are available to $\mathbf{O}$.  The physical interpretation of $[\mathcal{A}^{\mathit{i}}_{\mathit{jk}}, \mathcal{A}^{\mathit{i}}_{\mathit{lm}}] \neq \mathrm{0}$ for POVM components $\mathcal{A}^{\mathit{i}}_{\mathit{jk}}$ and $\mathcal{A}^{\mathit{i}}_{\mathit{lm}}$ must, therefore, also be realist, objectivist, and independent of the descriptions available to $\mathbf{O}$.  Suppose that at $t$, $\mathbf{C}$ is in a fine-grained state $|\mathbf{C}\rangle$ such the action $\mathcal{A}^{\mathit{i}}_{\mathit{jk}} |\mathbf{C}\rangle$ would cause $\mathbf{O}$ to record a value $a^{\mathit{i}}_{\mathit{jk}}$ and the action $\mathcal{A}^{\mathit{i}}_{\mathit{lm}} |\mathbf{C}\rangle$ would cause $\mathbf{O}$ to record a value $a^{\mathit{i}}_{\mathit{lm}}$; $|\mathbf{C}\rangle$ at $t$ is thus in the intersection $Im^{-1}(a^{\mathit{i}}_{\mathit{jk}}) \cap Im^{-1}(a^{\mathit{i}}_{\mathit{lm}})$ of the inverse images of $a^{\mathit{i}}_{\mathit{jk}}$ and $a^{\mathit{i}}_{\mathit{lm}}$.  In this case, the failure of commutativity can be expressed intuitively (e.g. \cite{schloss07} Ch. 2) in terms of the physical dynamics $\mathcal{H}$ by a pair of counterfactual conditionals:

\begin{quote}
If $|\mathbf{C}\rangle \in \mathit{Im}^{-1}(a^{\mathit{i}}_{\mathit{jk}}) \cap \mathit{Im}^{-1}(a^{\mathit{i}}_{\mathit{lm}})$ at $t$ and $\mathbf{O}$ does nothing at $t$, then at a subsequent $t + \Delta t$, $\mathcal{H} |\mathbf{C}\rangle \in \mathit{Im}^{-1}(a^{\mathit{i}}_{\mathit{jk}}) \cap \mathit{Im}^{-1}(a^{\mathit{i}}_{\mathit{lm}})$; however, if $|\mathbf{C}\rangle \in \mathit{Im}^{-1}(a^{\mathit{i}}_{\mathit{jk}}) \cap \mathit{Im}^{-1}(a^{\mathit{i}}_{\mathit{lm}})$ at $t$ and $\mathbf{O}$ measures either $\mathcal{A}^{\mathit{i}}_{\mathit{jk}}$ or  $\mathcal{A}^{\mathit{i}}_{\mathit{lm}}$ at $t$, then at a subsequent $t+ \Delta t$, $\mathcal{H} |\mathbf{C}\rangle \notin \mathit{Im}^{-1}(a^{\mathit{i}}_{\mathit{jk}}) \cap \mathit{Im}^{-1}(a^{\mathit{i}}_{\mathit{lm}})$. 
\end{quote}
Figure 5 illustrates this situation, the familiar ``dependence of the physical dynamics on the act of observation'' mentioned in the Introduction.

\psset{xunit=1cm,yunit=1cm}
\begin{pspicture}(0,0)(16,8)
\put(.5,7){(a)}
\put(9,7){(b)}
\put(2.4,6.7){$t$}
\pscircle(2.5,5){1.5}
\pscircle(2.5,3.5){1.5}
\put(1.5,5.5){$Im^{-1}(a^{\mathit{i}}_{\mathit{jk}})$}
\put(2.3,4.2){$|\mathbf{C}\rangle$}
\put(1.5,2.8){$Im^{-1}(a^{\mathit{i}}_{\mathit{lm}})$}
\put(4,4.25){\vector(1,0){1.5}}
\put(6.5,6.7){$t + \Delta t$}
\pscircle(7,5){1.5}
\pscircle(7,3.5){1.5}
\put(6,5.5){$Im^{-1}(a^{\mathit{i}}_{\mathit{jk}})$}
\put(6.5,4.2){$\mathcal{H} |\mathbf{C}\rangle$}
\put(6,2.8){$Im^{-1}(a^{\mathit{i}}_{\mathit{lm}})$}
\put(10.9,6.7){$t$}
\pscircle(11,5){1.5}
\pscircle(11,3.5){1.5}
\put(10,5.5){$Im^{-1}(a^{\mathit{i}}_{\mathit{jk}})$}
\put(10.8,4.2){$|\mathbf{C}\rangle$}
\put(10,2.8){$Im^{-1}(a^{\mathit{i}}_{\mathit{lm}})$}
\put(12.5,4.25){\vector(1,2){1}}
\put(12.5,4.25){\vector(2,1){1.3}}
\put(12.5,4.25){\vector(2,-1){1.3}}
\put(12.5,4.25){\vector(1,-2){1}}
\put(15,6.7){$t + \Delta t$}
\pscircle(15.5,5){1.5}
\pscircle(15.5,3.5){1.5}
\put(14.5,5.6){$Im^{-1}(a^{\mathit{i}}_{\mathit{jk}})$}
\put(13.5,6.4){$\mathcal{H} |\mathbf{C}\rangle$?}
\put(14.1,5.1){$\mathcal{H} |\mathbf{C}\rangle$?}
\put(14.1,3.2){$\mathcal{H} |\mathbf{C}\rangle$?}
\put(13.5,1.8){$\mathcal{H} |\mathbf{C}\rangle$?}
\put(14.5,2.7){$Im^{-1}(a^{\mathit{i}}_{\mathit{lm}})$}
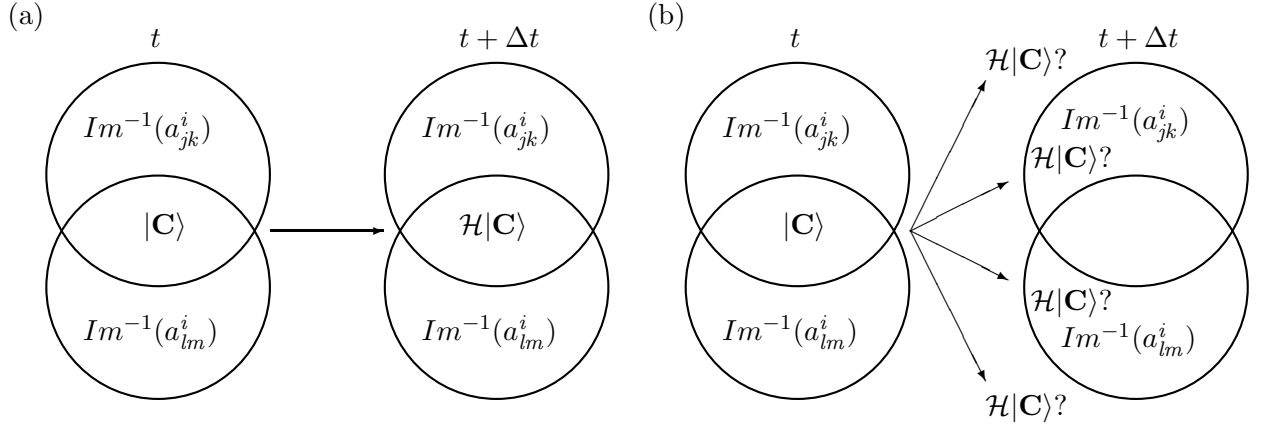
\put(0.5,1){\textit{Fig. 5: Dynamic evolution of $|\mathbf{C}\rangle$ without (a) and with (b) $\mathbf{O}$'s measurement of $\mathcal{A}^{\mathit{i}}_{\mathit{jk}}$ or  $\mathcal{A}^{\mathit{i}}_{\mathit{lm}}$ at $t$.}}
\put(1.7,.3){\textit{Part (b) shows the four possible post-measurement locations of $\mathcal{H} |\mathbf{C}\rangle$ if $[\mathcal{A}^{\mathit{i}}_{\mathit{jk}}, \mathcal{A}^{\mathit{i}}_{\mathit{lm}}] \neq \mathit{0}$.}}
\end{pspicture}

Implicit in this intuitive formulation of non-commutativity as a counterfactual conditional, and in Fig. 5a, is the idea that the observer could ``do nothing'' at $t$, thus avoiding the ``perturbation'' of $|\mathbf{C}\rangle$ with either $\mathcal{A}^{\mathit{i}}_{\mathit{jk}}$ or  $\mathcal{A}^{\mathit{i}}_{\mathit{lm}}$.  The definition of a minimal observer, however, permits $\mathbf{O}$ to ``do nothing'' only if the control values $s^{\mathit{i}}_{\mathrm{1}}, ..., s^{\mathit{i}}_{n^{\mathit{i}}}$ are not accepted, i.e. only if (to use the usual language of external systems momentarily) the ``system'' $\mathbf{S}^{\mathit{i}}$ is not identified as ``ready'' by the POVM $\lbrace \mathcal{S}^{\mathit{i}}_{\mathit{k}} \rbrace$.  If $\mathbf{S}^{\mathit{i}}$ is identified by the action of $\lbrace \mathcal{S}^{\mathit{i}}_{\mathit{k}} \rbrace$ as ready, $\mathbf{O}$ deterministically makes an observation and records a value.  The dynamics depicted in Fig. 5a is thus inconsistent with the condition that $|\mathbf{C}\rangle \in \mathit{Im^{-1}} \lbrace \mathcal{S}^{\mathit{i}}_{\mathit{k}} \rbrace$ at $t$.  Consistency with $|\mathbf{C}\rangle \in \mathit{Im^{-1}} \lbrace \mathcal{S}^{\mathit{i}}_{\mathit{k}} \rbrace$ at $t$ requires that if $|\mathbf{C}\rangle \in \mathit{Im}^{-1}(a^{\mathit{i}}_{\mathit{jk}}) \cap \mathit{Im}^{-1}(a^{\mathit{i}}_{\mathit{lm}})$ at $t$, $|\mathbf{C}\rangle \notin \mathit{Im}^{-1}(a^{\mathit{i}}_{\mathit{jk}}) \cap \mathit{Im}^{-1}(a^{\mathit{i}}_{\mathit{lm}})$ at an immediately-previous $t - \Delta t$.  This consistent situation is illustrated in Fig. 6, in which the uncertainties about the state of $\mathbf{C}$ before and after $t$ are symmetric.

\psset{xunit=1cm,yunit=1cm}
\begin{pspicture}(0,0)(16,8)
\put(2.5,5.7){$Im^{-1}(a^{\mathit{i}}_{\mathit{jk}})$}
\put(4.2,6.5){$\mathcal{H}^{\mathit{-1}} |\mathbf{C}\rangle$?}
\put(3.3,5.1){$\mathcal{H}^{\mathit{-1}} |\mathbf{C}\rangle$?}
\put(3.3,3.2){$\mathcal{H}^{\mathit{-1}} |\mathbf{C}\rangle$?}
\put(4.2,1.8){$\mathcal{H}^{\mathit{-1}} |\mathbf{C}\rangle$?}
\put(2.5,2.7){$Im^{-1}(a^{\mathit{i}}_{\mathit{lm}})$}
\pscircle(3.5,5.1){1.5}
\pscircle(3.5,3.5){1.5}
\put(2.9,6.7){$t - \Delta t$}
\put(6.5,4.25){\vector(-1,2){1}}
\put(6.5,4.25){\vector(-2,1){1.3}}
\put(6.5,4.25){\vector(-2,-1){1.3}}
\put(6.5,4.25){\vector(-1,-2){1}}
\put(7.9,6.7){$t$}
\pscircle(8,5){1.5}
\pscircle(8,3.5){1.5}
\put(7,5.5){$Im^{-1}(a^{\mathit{i}}_{\mathit{jk}})$}
\put(7.8,4.2){$|\mathbf{C}\rangle$}
\put(7,2.8){$Im^{-1}(a^{\mathit{i}}_{\mathit{lm}})$}
\put(9.5,4.25){\vector(1,2){1}}
\put(9.5,4.25){\vector(2,1){1.3}}
\put(9.5,4.25){\vector(2,-1){1.3}}
\put(9.5,4.25){\vector(1,-2){1}}
\put(12,6.7){$t + \Delta t$}
\pscircle(12.5,5){1.5}
\pscircle(12.5,3.5){1.5}
\put(11.5,5.6){$Im^{-1}(a^{\mathit{i}}_{\mathit{jk}})$}
\put(10.5,6.4){$\mathcal{H} |\mathbf{C}\rangle$?}
\put(11.1,5.1){$\mathcal{H} |\mathbf{C}\rangle$?}
\put(11.1,3.2){$\mathcal{H} |\mathbf{C}\rangle$?}
\put(10.5,1.8){$\mathcal{H} |\mathbf{C}\rangle$?}
\put(11.5,2.7){$Im^{-1}(a^{\mathit{i}}_{\mathit{lm}})$}

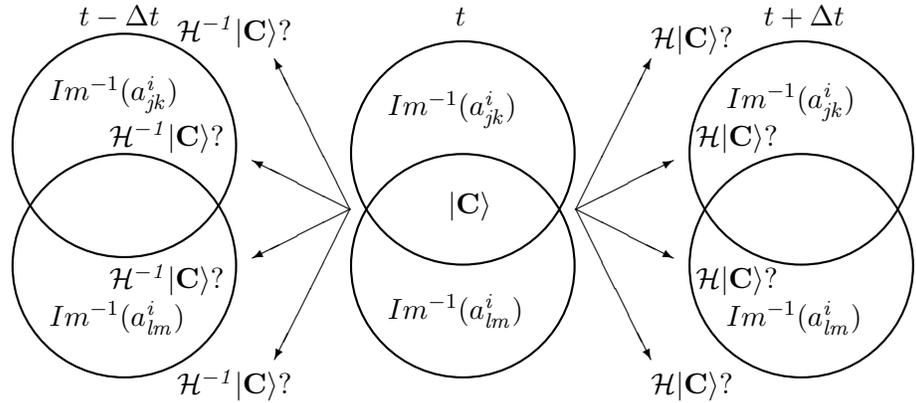
\put(1,.6){\textit{Fig. 6: Dynamic evolution of $|\mathbf{C}\rangle$ that is consistent at all times with $|\mathbf{C}\rangle \in \mathit{Im^{-1}} \lbrace \mathcal{S}^{\mathit{i}}_{\mathit{k}} \rbrace$ at $t$.}}
\end{pspicture}

Realism and objectivism demand that the forward and reverse dynamics of $\mathcal{H}$ depicted in Fig. 6 receive the same physical interpretation.  Viewing the dynamics symmetrically and considering $\mathbf{O}$'s control structure as shown in Fig. 4 makes the causal structure of the sequence from $t - \Delta t$ to $t$ clear: \textit{if} the physical evolution of $\mathbf{O \otimes C}$ under the action of $\mathcal{H}$ results in $|\mathbf{C}\rangle \in \mathit{Im}^{-1}(a^{\mathit{i}}_{\mathit{jk}}) \cap \mathit{Im}^{-1}(a^{\mathit{i}}_{\mathit{lm}})$ at $t$, \textit{either} $\mathcal{A}^{\mathit{i}}_{\mathit{jk}}$ or $\mathcal{A}^{\mathit{i}}_{\mathit{lm}}$ will be executed by $\mathbf{O}$ at $t$, with precedence determined by $\mathbf{O}$'s control structure.  The control structure of $\mathbf{O}$, however, is a \textit{virtual machine} implemented by the collection of physical degrees of freedom $\mathbf{O}$, the time evolution of which are driven by $\mathcal{H}$.  Every action of $\mathbf{O}$, therefore, is fully determined by $\mathcal{H}$ via the emulation mapping that defines $\mathbf{O}$ as a physically-implemented virtual machine.  Far from ``dependence of the physical dynamics on the act of observation,'' the transition from $t - \Delta t$ to $t$ illustrates the \textit{deterministic} dependence of the act of observation on the physical dynamics.  If the dynamics determines the observation from $t - \Delta t$ to $t$, however, it must determine the observation from $t$ to $t + \Delta t$ as well.  There is nothing particular to quantum mechanics in this claim: once information is viewed as physical, the conclusion that an interaction that transfers information from $\mathbf{C}$ to $\mathbf{O}$ also transfers information from $\mathbf{O}$ back to $\mathbf{C}$ follows straightforwardly from Newton's Third Law.

Given this physical interpretation of non-commutativity as a consequence of the reaction of $\mathbf{O}$ on $\mathbf{C}$ that is required by a time-symmetric, deterministic $\mathcal{H}$, $\mathbf{O}$ will observe non-commutativity between any pair of POVMs $\lbrace \mathcal{A}^{\mathit{i}}_{\mathit{jk}} \rbrace$ and $\lbrace \mathcal{A}^{\mathit{i}}_{\mathit{lm}} \rbrace$ with $j \neq l$ for which the action of $\mathcal{H}$ on $Im^{-1}(a^{\mathit{i}}_{\mathit{jk}})$ alters the subsequent distribution of degrees of freedom into $Im^{-1}(a^{\mathit{i}}_{\mathit{lm}})$ for some $m$ or vice versa.  Commutativity of $\lbrace \mathcal{A}^{\mathit{i}}_{\mathit{jk}} \rbrace$ and $\lbrace \mathcal{A}^{\mathit{i}}_{\mathit{lm}} \rbrace$ thus requires that $Im^{-1}(a^{\mathit{i}}_{\mathit{jk}})$ and $Im^{-1}(a^{\mathit{i}}_{\mathit{lm}})$ are separable under the dynamics $\mathcal{H}$ for all $k$ and $m$.  Operators that jointly measure the action of $\mathcal{H}$, in particular, will never satisfy this condition; hence such operators cannot commute.  It is impossible, moreover, for any minimal observer to predict the effect of $\mathcal{H}$ on a given $Im^{-1}(a^{\mathit{i}}_{\mathit{jk}})$ and alter the choice of subsequent measurement to avoid the appearance of non-commutativity, as doing so would require an ability to represent the state of $\mathbf{O \otimes C}$, a state about which minimal observers are objectively ignorant.

The present framework offers, therefore, a straightforward answer to van Fraassen's \cite{vanFraassen91} question ``How could the world possibly be the way a quantum theory says it is?''  The world is a physically-implemented information channel, it evolves through the action of a time-symmetric, deterministic dynamics that satisfies decompositional equivalence and counterfactual definiteness, and it contains minimal observers implementing pairs of POVMs with non-separable inverse images, in particular pairs of POVMs that jointly measure action.  Within the present framework, the more interesting question is the reverse of van Fraassen's: what would the world have to be like for \textit{classical} mechanics to be true, i.e. for dynamics to be time-symmetric, deterministic, satisfy decompositional equivalence and counterfactual definiteness, and for all possible physical observables to commute?  There are two answers.  First, the world would be classical if information transfer required zero time.  If information could be transferred instantaneously, multiple POVMs could act on a single channel state $|\mathbf{C}\rangle$ without intervening reactions of $\mathbf{O}$ on $\mathbf{C}$.  Second, the world would be classical if observers had effectively infinite coding capacity.  With infinite coding capacity, observers could in principle realize the Laplacian dream of completely modeling $\mathcal{H}$, and hence designing time-dependent POVMs with inverse images that accurately predicted the trajectory from any $|\mathbf{C}\rangle$ to the unique subsequent $\mathcal{H} |\mathbf{C}\rangle$.  These conditions could both be true if information was not physical.  Hence the operator commutativity required by classical mechanics could be true if information were not physical, and can be derived given a fundamental assumption that information is not physical, that information processing in principle costs nothing, is free (c.f. \cite{landauer99} where free information is identified with classicality).  What the empirical success of quantum mechanics tells us is that information \textit{is} physical: that information processing is \textit{not} free.

\section{Physical Interpretation of Bell's Theorem, the Born Rule and Decoherence}

The previous section showed that, given reasonable, traditional, and not explicitly quantum-mechanical assumptions about the dynamics driving the evolution of a physical information channel, any physically-implemented minimal observer equipped with sufficiently high-resolution POVMs will discover one of the primary features of the quantum world: pairs of POVMs with mutually non-separable inverse images, including pairs of POVMs that jointly measure action, will not commute.  This section will show that minimal observers equipped with sufficiently high-resolution POVMs will also discover several other canonically ``quantum'' phenomena.  Before proceeding, however, it is useful to summarize, in Table 1, the meanings given to the fundamental terms of the standard quantum-mechanical formalism by the formal framework for describing the $\mathbf{O - C}$ interaction developed in the last two sections.  

\psset{xunit=1cm,yunit=1cm}
\begin{pspicture}(0,0)(16,8)
\pspolygon(2,7)(14,7)(14,1.4)(2,1.4)
\put(2.4,6.3){\textbf{Standard quantum formalism}}
\put(9.3,6.3){\textbf{Current framework}}
\put(2,6){\line(1,0){12}}
\put(2.3,5.3){Quantum system $\mathbf{S}$, a collection}
\put(2.7,4.7){of degrees of freedom}
\put(8.3,5.3){$Im^{-1}(\lbrace \mathcal{S}^{\mathit{i}}_{\mathit{k}} \rbrace)$, the (non-NULL)}
\put(8.6,4.7){inverse image in $\mathbf{C}$ of a POVM}
\put(2,4.4){\line(1,0){12}}
\put(2.3,3.8){Quantum state $|\mathbf{S}\rangle$ at $t$}
\put(8.3,3.8){$Im^{-1}(a_{jk})$ in $\mathbf{C}$ at $t$ for value $a_{jk}$}
\put(8.6,3.2){of a POVM component $\mathcal{A}^{\mathit{i}}_{\mathit{jk}}$}
\put(2,3){\line(1,0){12}}
\put(2.3,2.4){Observable $\mathcal{A}$, defined over}
\put (2.7,1.8){states of any quantum system}
\put(8.3,2.4){$\lbrace \mathcal{A}^{\mathit{1}}_{\mathit{jk}} \rbrace ... \lbrace \mathcal{A}^{\mathit{N}}_{\mathit{jk}} \rbrace$, a set of POVMs}
\put(8.6,1.8){defined over states of $\mathbf{C}$}
\put(8,1.4){\line(0,1){5.5}}
\put(.1,.5){\textit{Table 1: Meanings assigned to terms in the standard quantum formalism by the current framework.}}
\end{pspicture}

As shown in Table 1, the fundamental difference between the current framework and the standard quantum formalism is the meaning assigned to the notion of a quantum system.  In the standard quantum formalism, a quantum system is a collection of physical degrees of freedom, and any quantum system is observable in principle.  In the current framework, an observable quantum system is the non-NULL inverse image, in a physical channel $\mathbf{C}$, of a physically-implemented POVM with a finite number of finite, real output values.  The current framework thus \textit{limits} quantum theory by placing an observer-relative, information-theoretic restriction on what ``counts'' as an observable quantum system: the POVM $\lbrace \mathcal{S}^{\mathit{i}}_{\mathit{k}} \rbrace$ must be physically implemented by an observer $\mathbf{O}$ in order for the ``quantum system'' it detects to exist for $\mathbf{O}$.  Thus in the current framework, to paraphrase Fuchs' \cite{fuchs10} paraphrase of de Finetti, ``quantum systems do not exist'' as objective, ``given'' entities.  This does not, clearly, mean that the \textit{stuff} composing quantum systems does not exist; both $\mathbf{C}$ and $\mathbf{O}$ are implemented by \textit{physical} degrees of freedom.  What it means is that their \textit{boundaries} do not exist.  Systems are defined only by observer-imposed decompositions, and physical dynamics do not respect decompositional boundaries.

Quantum states are, in the current framework, equivalence classes under the components of a POVM $\lbrace \mathcal{A}^{\mathit{i}}_{\mathit{jk}} \rbrace$ of states of $\mathbf{C}$ that are indistinguishable, in principle, by an observer implementing $\lbrace \mathcal{A}^{\mathit{i}}_{\mathit{jk}} \rbrace$.  As discussed in Sect. 3, other than whether $|\mathbf{C}\rangle$ is identified by an available POVM $\lbrace \mathcal{S}^{\mathit{i}}_{\mathit{k}} \rbrace$ and the values $a_{jk}$ assigned by the $\lbrace \mathcal{P}^{\mathit{i}}_{\mathit{k}} \rbrace$-selected $j^{th}$ available observable $\lbrace \mathcal{A}^{\mathit{i}}_{\mathit{jk}} \rbrace$ that are obtained in the course of a finite sequence of measurements, observers in the current framework are \textit{objectively ignorant} about quantum states.  No physical state $|\mathbf{C}\rangle$ of the channel, and therefore no physical state of any ``system'' $\mathbf{S}$, can be either fully characterized or demonstrated to be replicated by any minimal observer, regardless of the amount of data that observer collects.  A world in which no observer is able, in principle, to identify any quantum state as a replicate of any other quantum state is, however, equivalent from the perspective of such an observer to a world in which quantum states cannot be replicated.  The observational consequences of objective ignorance regarding the replication of quantum states are, therefore, equivalent to the observational consequences of the no-cloning theorem \cite{wooters82}, which forbids the replication of unknown quantum states.  These consequences are realized objectively in the current framework for \textit{all} quantum states, since all are ``unknown'' to all observers.  In the current framework, the effective inability to clone quantum states is a consequence of the physicality of information and the boundarylessness of quantum systems defined as inverse images of POVMs.

In the current framework, no-cloning renders all observational results observer-specific.  Any two observers $\mathbf{O}$ and $\mathbf{O}^{\prime}$ are objectively ignorant about whether the inverse images of any two POVMs $\lbrace \mathcal{A}^{\mathbf{O}\mathit{i}}_{\mathit{jk}} \rbrace$ and $\lbrace \mathcal{A}^{\mathbf{O}^{\prime} \mathit{i}}_{\mathit{lm}} \rbrace$ are the same subsets of $\mathbf{C}$, whether these POVMs commute or not.  Whether two observers share observables can, therefore, at best be established ``for all practical purposes'' by comparing the results of multiple observations.  Hence it cannot be assumed, without qualifications, that two distinct observers have both measured a single observable such as $\hat{x}$ for a single system $\mathbf{S}$.  This reflects laboratory reality: whether an observation has been successfully replicated in all details is always subject to question.

With these understandings of the familiar terms, the physical meaning of Bell's theorem \cite{bell64} for a minimal observer becomes clear.  Consider an observer who measures the same observable on two different ``systems'' $\mathbf{S}^{\mathrm{1}}$ and $\mathbf{S}^{\mathrm{2}}$ employing triples $(\lbrace \mathcal{S}^{\mathrm{1}}_{\mathit{k}} \rbrace, \lbrace \mathcal{P}^{\mathrm{1}}_{\mathit{k}} \rbrace, \lbrace \mathcal{A}^{\mathrm{1}}_{\mathit{jk}} \rbrace)$ and $(\lbrace \mathcal{S}^{\mathrm{2}}_{\mathit{k}} \rbrace, \lbrace \mathcal{P}^{\mathrm{2}}_{\mathit{k}} \rbrace, \lbrace \mathcal{A}^{\mathrm{2}}_{\mathit{jk}} \rbrace)$ of POVMs at times $t$ and $t + \Delta t$ respectively.  Between $t$ and $t + \Delta t$, the state of $\mathbf{C}$ evolves from $|\mathbf{C}\rangle$ to $\mathcal{H} |\mathbf{C}\rangle$.  Clearly $|\mathbf{C}\rangle \in \mathit{Im}^{-1}(\lbrace \mathcal{S}^{\mathrm{1}}_{\mathit{k}} \rbrace)$ at $t$ and $\mathcal{H} |\mathbf{C}\rangle \in \mathit{Im}^{-1}(\lbrace \mathcal{S}^{\mathrm{2}}_{\mathit{k}} \rbrace)$ at $t + \Delta t$; otherwise the measurements could not be performed.  What is relevant to Bell's theorem is whether these inverse images overlap, and in particular, whether $Im^{-1}(\lbrace \mathcal{A}^{\mathrm{1}}_{\mathit{jk}} \rbrace)$ evaluated at $t$ intersects $\mathit{Im}^{-1}(\lbrace \mathcal{A}^{\mathrm{2}}_{\mathit{lm}})$ evaluated at $t + \Delta t$ for any $j$ and $l$.  If this intersection is empty, the measured ``states'' $|\mathbf{S}^{\mathrm{1}}\rangle$ and $|\mathbf{S}^{\mathrm{2}}\rangle$ are separable.  However, the intersection of the inverse image $Im^{-1}(\lbrace \mathcal{A}^{\mathrm{1}}_{\mathit{jk}} \rbrace)$ at $t$ and the inverse image $\mathit{Im}^{-1}(\lbrace \mathcal{A}^{\mathrm{2}}_{\mathit{jk}} \rbrace)$ at $t + \Delta t$ is only guaranteed to be empty if $\mathcal{H}$ respects the $\mathbf{S}^{\mathrm{1}}$ - $\mathbf{S}^{\mathrm{2}}$ boundary, and \textit{assuming} that $\mathcal{H}$ respects the $\mathbf{S}^{\mathrm{1}}$ - $\mathbf{S}^{\mathrm{2}}$ boundary violates decompositional equivalence.  Therefore, the default assumption must be that $Im^{-1}(\lbrace \mathcal{A}^{\mathrm{1}}_{\mathit{jk}} \rbrace)$ at $t$ may overlap $\mathit{Im}^{-1}(\lbrace \mathcal{A}^{\mathrm{2}}_{\mathit{jk}} \rbrace)$ at $t + \Delta t$, and hence that $|\mathbf{S}^{\mathrm{1}}\rangle$ and $|\mathbf{S}^{\mathrm{2}}\rangle$ cannot be regarded as separable.  That separability between apparently-distinct systems cannot be assumed by default is the operational content of Bell's theorem, accepting the horn of the dilemma on which counterfactual definiteness and hence the ability to talk about the inverse images of POVMs is assumed.

The problem with the classical reasoning that produces Bell's inequality, on the current framework, is that it assumes that observers can have perfect information about distant systems.  If $\mathbf{O}$ is making a local measurement of $\mathbf{S}^{\mathrm{1}}$ at $t$, and $\mathbf{S}^{\mathrm{1}}$ has a spacelike separation from $\mathbf{S}^{\mathrm{2}}$ at $t$, then $\mathbf{O}$ cannot be making a local measurement of $\mathbf{S}^{\mathrm{2}}$ at $t$.  If at some later time $t + \Delta t$ $\mathbf{O}$ writes down a joint probability distribution for particular states $|\mathbf{S}^{\mathrm{1}}\rangle$ and $|\mathbf{S}^{\mathrm{2}}\rangle$ at $t$, $\mathbf{O}$ must be in possession at $t + \Delta t$ of data obtained about $|\mathbf{S}^{\mathrm{2}}\rangle$ at $t$, such as a report of the state of $\mathbf{S}^{\mathrm{2}}$ at $t$ from some other observer, e.g. Alice, that is was local to $\mathbf{S}^{\mathrm{2}}$ at $t$.  The delivery of this report from Alice to $\mathbf{O}$ requires a physical channel, with which $\mathbf{O}$ must interact, using an appropriate POVM, in order to extract the information contained in the report.  Writing down the joint probability distribution for $|\mathbf{S}^{\mathrm{1}}\rangle$ and $|\mathbf{S}^{\mathrm{2}}\rangle$ at $t$ therefore requires that $\mathbf{O}$ make two local measurements, one of $|\mathbf{S}^{\mathrm{1}}\rangle$ at $t$, and one of the report from Alice at the later time $t + \Delta t$.  Only if the inverse images of the two POVMs required to make these two measurements are separable is the classical assumption of perfect information transfer from Alice to $\mathbf{O}$ warranted.  In standard quantum-mechanical practice, $\mathbf{O}$'s interactions with a macroscopic Alice at $t + \Delta t$ are assumed to separable from Alice's interactions with $\mathbf{S}^{\mathrm{2}}$ at $t$ due to decoherence; any entanglement between Alice and $\mathbf{S}^{\mathrm{2}}$ is assumed to be lost to the environment in a way that renders it inaccessible to $\mathbf{O}$.  This assumption, however, rests on an implicit assumption that $\mathbf{O}$ can distinguish Alice from the background of the environment without making a measurement of Alice's state \cite{fields11a}, e.g. before asking for her report.  If Alice is microscopic - for example, if Alice is a single photon - this latter assumption is unwarranted, as is the assumption that Alice is no longer entangled with $\mathbf{S}^{\mathrm{2}}$ at $t + \Delta t$.  A minimal observer, however, cannot identify any system other than by making a measurement of that system's state.  A minimal observer cannot, therefore, assume that decoherence has dissipated any previous entanglement into the environment; as will be described below, for a minimal observer decoherence is a property of information channels, not an observer-independent property of system-environment interactions.  Hence as discussed above, a minimal observer cannot assume that the inverse images of any two POVMs are separable; for a minimal observer, the default assumption must be that any two systems are entangled.  A minimal observer cannot, therefore, assume perfect information transfer from a distant source of data, and hence cannot derive Bell's inequality for spacelike separated systems using classical conditional probabilities that assume perfect information transfer.  For a minimal observer, therefore, the failure of Bell's inequality is expected, and the prediction of its failure by minimal quantum mechanics is positive evidence for the theory's correctness.

Viewing both quantum systems and quantum states as inverse images of POVMs also enables a straightforward physical interpretation of the Born rule.  Observers are objectively ignorant, at all times, of both the state $|\mathbf{C}\rangle$ of the information channel and the dynamics $\mathcal{H}$ driving its evolution.  By assuming decompositional equivalence, however, an observer can be confident that the future evolution of $|\mathbf{C}\rangle$ will not depend on the locations or boundaries within the state space of $\mathbf{C}$ of the inverse images of the POVMs $\lbrace \mathcal{S}^{\mathit{i}}_{\mathit{k}} \rbrace$, $\lbrace \mathcal{P}^{\mathit{i}}_{\mathit{k}} \rbrace$ or $\lbrace \mathcal{A}^{\mathit{i}}_{\mathit{jk}} \rbrace$.  Such an observer can, therefore, be confident that the probability of obtaining an outcome $a^{\mathit{i}}_{\mathit{jk}}$ following a successful application of $\lbrace \mathcal{S}^{\mathit{i}}_{\mathit{k}} \rbrace$, $\lbrace \mathcal{P}^{\mathit{i}}_{\mathit{k}} \rbrace$ and $\lbrace \mathcal{A}^{\mathit{i}}_{\mathit{jk}} \rbrace$ to $|\mathbf{C}\rangle$ at some future time $t$ will depend only on the number of physical states within $Im^{-1}(a^{\mathit{i}}_{\mathit{jk}})$ relative to the total number of states within of $Im^{-1}(\lbrace \mathcal{S}^{\mathit{i}}_{\mathit{k}} \rbrace)$ at $t$.  The Born rule expresses this confidence that $\mathcal{H}$ respects decompositional equivalence.

Let $P(a^{\mathit{i}}_{\mathit{jk}} | ij, t)$ be the probability that $\mathbf{O}$ records the value $a^{\mathit{i}}_{\mathit{jk}}$ at some future time $t$ given that $\mathbf{O}$ has, immediately prior to $t$, identified a ``system'' $\mathbf{S}^{\mathit{i}}$ by successful application of $\lbrace \mathcal{S}^{\mathit{i}}_{\mathit{k}} \rbrace$ and selected an observable $\lbrace \mathcal{A}^{\mathit{i}}_{\mathit{jk}} \rbrace$ by successful application of $\lbrace \mathcal{P}^{\mathit{i}}_{\mathit{k}} \rbrace$.  Given these conditions, $\mathbf{O}$ deterministically records some value $a^{i}_{jk}$, so $\sum_k P(a^{\mathit{i}}_{\mathit{jk}} | ij, t) = 1$.  If the POVM $\lbrace \mathcal{A}^{\mathit{i}}_{\mathit{jk}} \rbrace$ is restricted to only the components with $k \neq 0$ and hence considered to act only on the subspace $Im^{-1} \lbrace \mathcal{A}^{\mathit{i}}_{\mathit{jk}} \rbrace$ of $\mathbf{H}_{\mathbf{C}}$, it can be renormalized so that $\sum_{k} \mathcal{A}^{\mathit{i}}_{\mathit{jk}}$ is the Identity on $Im^{-1} \lbrace \mathcal{A}^{\mathit{i}}_{\mathit{jk}} \rbrace$.  Following the notation used by Zurek in his proof of the Born rule from envariance \cite{zurek05env}, let $m_{\mathit{k}}$ be the number of states in $Im^{-1}(a^{\mathit{i}}_{\mathit{jk}})$ and $M = \sum_{\mathit{k}} m_{\mathit{k}}$ be the number of states in $Im^{-1} \lbrace \mathcal{A}^{\mathit{i}}_{\mathit{jk}} \rbrace$; $Im^{-1} \lbrace \mathcal{A}^{\mathit{i}}_{\mathit{jk}} \rbrace$ then corresponds to the ``counter'' ancilla $C$ in Zurek's proof, each of the $k$ components of which contains $m_{\mathit{k}}$ fine-grained states.  What Zurek shows is that (in his notation \cite{zurek05env} but suppressing phases) if a joint system-environment state $|\psi_{SE}\rangle$ has a Schmidt decomposition $\sum_{k=1}^{N} a_{\mathit{k}} |s_{\mathit{k}}\rangle |e_{\mathit{k}}\rangle$ with $a_{\mathit{k}} \propto \sqrt{m_{\mathit{k}}}$, an ancilla $C$ of $M$ fine-grained states can be chosen with $k$ mutually-orthogonal components $C_{\mathit{k}}$ such that $C = \cup_{\mathit{k}} C_{\mathit{k}}$ and each $C_{\mathit{k}}$ contains $m_{\mathit{k}}$ fine-grained states.  Using the $C_{\mathit{k}}$ to count the number of fine-grained states available for entanglement with any given joint state $|s_{\mathit{k}}\rangle |e_{\mathit{k}}\rangle$, Zurek then shows that the probability $p_{\mathit{k}}$ of observing $|s_{\mathit{k}}\rangle |e_{\mathit{k}}\rangle$ is $m_{\mathit{k}}/M$, which equals $|a_{\mathit{k}}|^{\mathrm{2}}$ by the definition of $C$, giving the Born rule.

In the present context, the formalism of Zurek's proof provides a constructive definition of the unknown future quantum state on which a POVM $\lbrace \mathcal{A}^{\mathit{i}}_{\mathit{jk}} \rbrace$ can act to produce $a^{k}_{\mathit{jk}}$ as a recorded outcome.  The inverse image $Im^{-1} \lbrace \mathcal{A}^{\mathit{i}}_{\mathit{jk}} \rbrace$ is the subset of $\mathbf{C}$ that ``encodes'' the ''quantum state'' of the ``system'' $\mathbf{S}^{\mathit{i}}$ picked out by the POVM $\lbrace \mathcal{S}^{\mathit{i}}_{\mathit{k}} \rbrace$; the rest of $\mathbf{C}$ (i.e. $\mathbf{C} \setminus \mathit{Im^{-1}} \lbrace \mathcal{S}^{\mathit{i}}_{\mathit{k}} \rbrace$) is the ``environment'' of $\mathbf{S}^{\mathit{i}}$.  Hence Zurek's ``$|\psi_{SE}\rangle$'' is a coarse-grained representation of $|\mathbf{C}\rangle$, where the coarse-grained basis vectors ``$|s_{\mathit{k}}\rangle$'' and ``$|e_{\mathit{k}}\rangle$'' span the subpaces $Im^{-1} \lbrace \mathcal{A}^{\mathit{i}}_{\mathit{jk}} \rbrace$ and $\mathbf{C} \setminus \mathit{Im^{-1}} \lbrace \mathcal{S}^{\mathit{i}}_{\mathit{k}} \rbrace$ respectively.  Given Zurek's assumption that all system states are measureable, the $|s_{\mathit{k}}\rangle$ can be readily identified as the $Im^{-1}(a^{\mathit{i}}_{\mathit{jk}})$ for the POVM $\lbrace \mathcal{A}^{\mathit{i}}_{\mathit{jk}} \rbrace$; the $|e_{\mathit{k}}\rangle$ are notional, as they are for Zurek.  Hence the physical content of the Born rule is that, given decompositional equivalence, the inverse images $Im^{-1}(a^{\mathit{i}}_{\mathit{jk}})$ can be regarded as coarse-grained basis vectors for $Im^{-1} \lbrace \mathcal{A}^{\mathit{i}}_{\mathit{jk}} \rbrace$ that together provide a complete specification of the state of $Im^{-1} \lbrace \mathcal{A}^{\mathit{i}}_{\mathit{jk}} \rbrace$ as measurable by $\mathbf{O}$.  This is in fact the role of the Born rule in standard quantum theory: it assures that the probabilities $P(a^{\mathit{i}}_{\mathit{jk}} | ij, t)$ are exhausted by the amplitudes (squared) of the measureable basis vectors $|s_{\mathit{k}}\rangle$ of the identified system of interest.

Interpreting the Born rule in this way provides, in turn, a natural physical interpretation of decoherence.  Observers, as noted in Sect. 3, are virtual machines implemented by physical degrees of freedom.  Any ``system'' identified by a POVM $\lbrace \mathcal{S}^{\mathit{i}}_{\mathit{k}} \rbrace$ implemented by an observer is, therefore, itself a virtual entity: ``quantum systems do not exist'' as objective entities.  Decoherence must, therefore, be a virtual process acting on the information available to an observer, not a physical process acting on the degrees of freedom that implement $\mathbf{C}$.  Representing decoherence in this way requires re-interpretating it as an intrinsic property of a (quantum) information channel.  Such a re-interpretation can be motivated by noting that the usual physical interpretation of decoherence relies on the identification of quantum systems over time and is therefore deeply circular \cite{fields10, fields11a}.

In standard quantum theory, decoherence occurs when a quantum system $\mathbf{S}$ is suddenly exposed to a surrounding environment $\mathbf{E}$.  The $\mathbf{S - E}$ interaction $\mathcal{H}_{\mathbf{S - E}}$ rapidly couples degrees of freedom of $\mathbf{S}$ to degrees of freedom of $\mathbf{E}$, creating an entangled joint state in which degrees of freedom ``of $\mathbf{S}$'' can no longer be distinguished from degrees of freedom ``of $\mathbf{E}$.''  The phase coherence of the previous pure state $|\mathbf{S}\rangle$ is dispersed into the entangled joint system $\mathbf{S - E}$.  Under ordinary circumstances decoherence is very fast; Schlosshauer (\cite{schloss07} Ch. 3) estimates decoherence times for macroscopic objects exposed to ambient photons and air pressure to be many orders of magnitude less than the light-transit times for such objects (e.g. $10^{-31}$ s to spatially decohere a $10^{-3}$ cm dust particle at normal air pressure versus a light-transit time of $10^{-14}$ s).  It is, therefore, safe to regard all ordinary macroscopic objects exposed to the ordinary macroscopic environment as fully decohered.

It is worth asking, however, what is meant physically by the supposition that $\mathbf{S}$ is ``suddenly exposed'' to $\mathbf{E}$.  If $\mathbf{S}$ is ``suddenly exposed'' to $\mathbf{E}$ at some time $t$, it must have been isolated from $\mathbf{E}$ before $t$.  Call ``$\mathbf{F}$'' whatever imposes the force required to isolate $\mathbf{S}$ from $\mathbf{E}$.  On pain of infinite regress, $\mathbf{F}$ must be in contact with $\mathbf{E}$, in which case decoherence theory tells us that $\mathbf{F}$ and $\mathbf{E}$ are almost instantaneously entangled.  The interaction of $\mathbf{F}$ with $\mathbf{S}$ that imposes the force that keeps $\mathbf{S}$ isolated will, however, also entangle $\mathbf{S}$ with $\mathbf{F}$.  Unless $\mathbf{F}$ can be partitioned into separable components $\mathbf{F1}$ and $\mathbf{F2}$ that separately interact with $\mathbf{S}$ and $\mathbf{E}$ respectively, however, neither $|\mathbf{F} \otimes \mathbf{S}\rangle$ nor $|\mathbf{F} \otimes \mathbf{E}\rangle$ can be considered to be pure states, and nothing prevents the spread of entanglement from $\mathbf{S}$ to $\mathbf{E}$.  Hence unless $\mathbf{F}$ can be partitioned into separable components, $\mathbf{S}$ has never been isolated, and can never be ``suddenly exposed.''  In practice, $\mathbf{F}$ is often a piece of laboratory apparatus such as an ion trap, that interacts with an ``isolated'' system on one surface and the environment on another.  The assumption that $\mathbf{F}$ can be partitioned into separable systems is, effectively, the assumption of an internal boundary within $\mathbf{F}$ that is not crossed by any entangling interactions.  Such an internal boundary would, however, ``isolate'' everything inside it, and hence require another internal boundary to enforce this isolation. Such an infinite regress of boundaries is impossible; hence no such boundary can exist.

That this reasoning applies across the dynamical domains defined by the relation between the self and interaction Hamiltonians of $\mathbf{S}$ (e.g. \cite{schloss07, landsman07}) can be seen by considering a high-energy cosmic ray that collides with the Earth.  During its transit of interplanetary space and the upper atmosphere, the interaction of the cosmic ray with its immediate environment is small; it can be considered ``isolated'' as long as no measurements of its state are made.  Its sudden collision with dense matter (e.g. a scintillation counter) ``exposes'' it to the local environment defined by that matter, a local environment that is contiguous with the larger environment of the universe as a whole.  This ``sudden exposure'' is, however, an artifact of the limited view of the cosmic ray's history just described.  The cosmic ray was produced by a nuclear reaction, e.g. in the Sun.  Prior to that reaction, its future components were fully exposed to the local environment of the Sun, a local environment that, like the dense matter on Earth, was contiguous with the larger environment of the universe as a whole.  The pre-reaction entanglement between components of the future cosmic ray and other components of the Sun, and hence with other components of the universe as a whole, is not physically destroyed by the formation and flight of the cosmic ray; it is merely inaccessible to observers on Earth, who are only able to experimentally take note of the later, local entanglement between the cosmic ray and the Earth-bound matter with which it collides.  It is widely acknowledged that the notion of an ``isolated system'' is a holdover from classical physics; Schlosshauer, for example, notes that ``the idealized and ubiquitous notion of isolated systems remained a guiding principle of physics and was adopted in quantum mechanics without much further scrutiny'' (\cite{schloss07} p. 1).  Yet if quantum systems are never isolated, if all physical degrees of freedom are entangled at all times with all other physical degrees of freedom, what is the physical meaning of decoherence?

Standard quantum theory resolves this paradox formally.  The formalism distinguishes $\mathbf{S}$ from $\mathbf{E}$ by giving them different names.  The representation $|\mathbf{S} \otimes \mathbf{E}\rangle = \mathit{\sum_{ij} \lambda_{ij} |s_{i}\rangle |e_{j}\rangle}$ of the entangled joint state preserves this distinction, as does the joint density $\rho = \frac{1}{2} \sum_{ij} |s_{i}\rangle \langle s_{j}||e_{i}\rangle \langle e_{j}|$ and its partial trace over $\mathbf{E}$, $\rho_{\mathbf{S}} = \frac{1}{N} \sum_{ij=1}^{N} |s_{i}\rangle \langle s_{j}| \langle e_{i}|e_{j}\rangle$.  These representations all assume, implicitly, that $\mathbf{S}$ can be identified against the background of $\mathbf{E}$; the partial trace additionally assumes, usually explicitly, that $\mathbf{O}$ is employing an observable $\mathcal{A \otimes I}$ that measures states of $\mathbf{S}$ in some basis but acts as the identity operator on states of $\mathbf{E}$.  It is this latter assumption that is expressed by the standard proviso that $\mathbf{O}$ cannot or does not observe the states of $\mathbf{E}$.  Given these assumptions, however, the claim that decoherence \textit{explains} $\mathbf{O}$'s ability to distinguish $\mathbf{S}$ from $\mathbf{E}$ by providing a physical mechanism for the ``emergence of classicality'' is clearly circular: the ``emergence'' is built in from the beginning by assigning the distinct \textit{names} $\mathbf{S}$ and $\mathbf{E}$ and assuming that they refer to different things.  Indeed, the role of decoherence in standard quantum theory appears to be that of an axiom, somewhat more subtle that von Neumann's axiom of wave-function collapse, stating that observers can distinguish quantum systems from their environments even though the two are always and inevitably entangled.  The statement ``decoherence is a physical process'' thus appears entirely equivalent to Zurek's ``axiom(o).''

To see how ``axiom(o)'' is employed in practice, consider the now-classic cavity-QED experiments of Brune \textit{et al.} \cite{brune96} (reviewed in \cite{schloss07} Ch. 6), in which decoherence of a mesoscopic ``Schr\"odinger cat'' created by coupling a well-defined excited state of a single Rb atom to a weak photon field inside a superconducting cavity is monitored as a function of time and experimental conditions.  In the standard language of quantum systems and states, the system $\mathbf{S}$ in this case provides two observables, the state $e$ (excited) or $g$ (ground) of an Rb atom after it has traversed the cavity, and the correlation $P_{ij}(\Delta t)$ between the states of successive atoms $i$ and $j$ arriving at the detector with a time difference of $\Delta t$.  The experimental outcomes are: (1) varying the coupling between the atomic state and the photon field varies the amount of information about the traversing atom's state that was stored in the field (\cite{brune96} Fig. 3); and (2) varying the time interval $\Delta t$ varies the amount of information about the $i^{th}$ atom's state that could be extracted from the $j^{th}$ atom's state (\cite{brune96} Fig. 5).  The first result demonstrates that increasing the local interaction between two \textit{identified} degrees of freedom (by increasing the coupling) increases the entanglement between \textit{those} degrees of freedom.  The second result demonstrates that after the local interaction between the two identified degrees of freedom (after the $i^{th}$ atom leaves the cavity), the entanglement between those degrees of freedom dissipates; the field in the cavity is also entangled with the atoms in the walls of the cavity, and this latter entanglement decoheres the ``information'' about the $i^{th}$ atom's state that ``the atom leaves in (the cavity) $C$'' (\cite{brune96} p. 4889).  Critical to this explanation is the tacit assumption that the states of the atoms in the walls of the cavity are not themselves observed, or equivalently, that the atoms in the walls of the cavity are themselves entangled with the general environment in which the apparatus is embedded.  But, this assumption comes with the implicit proviso that this prior system - environment entanglement \textit{does not prevent the identification of quantum states} of the individual Rb atoms traversing the cavity.  This assumption that the individual Rb atoms can be regarded ``objectively'' even in the presence of system - environment entanglement is an instance of ``axiom(o).''

The current framework alters this standard account of the physics by re-casting it in informational terms and rejecting the tacit assumption that the $i^{th}$ and $j^{th}$ Rb atoms are distinguishable quantum systems.  The ``system'' $\mathbf{S}^{\mathit{B}}$ in this framework (``$B$'' for Brune \textit{et al.}) is the inverse image of a POVM $\lbrace \mathcal{S}^{\mathit{B}}_{\mathit{k}} \rbrace$ with control variables $s^{B}_{\mathrm{1}}, ..., s^{B}_{n^{B}}$.  Distinct acceptable sets of values of these variables describe distinct preparation conditions for the system.  This system can be considered an information channel from $Im^{-1}(\lbrace \mathcal{A}^{\mathit{B}}_{\mathit{g}}, \mathcal{A}^{\mathit{B}}_{\mathit{e}} \rbrace)$ to $\mathbf{O}$, where the components of $\mathcal{A}^{\mathit{B}}$ report the outcomes $g$ and $e$ respectively.  In this representation, long-lived entanglement between the atom traversing the cavity and the photon field within it causes delocalization in time of the outcome: the values of the control variables $s^{B}_{\mathrm{1}}, ..., s^{B}_{n^{B}}$ - specifically, those indicating the mirror separation and hence tuning of the cavity - can be adjusted in a way that smears an outcome $g$ (for example) out over pairs of applications of $\mathcal{A}^{\mathit{B}}$.  Figure 7 illustrates this smearing in time using a simple circuit model, in which the (approximately) fixed ``resistance'' $R$ represents information loss from the channel (e.g. the approximately fixed coupling of the photon field to the cavity) and the variable ``capacitance'' $C$ represents the intrinsic memory of the channel (e.g. the manipulable coupling of the atomic beam to the photon field).  An instantaneous input impulse $\delta (t - t_{0})$ at $t = t_{0}$ results in an output $\propto e^{-t/RC}$ for $t > t_{0}$ at $\mathbf{O}$.  The time constant $RC$ is the decoherence time; it is a measure of the channel's memory of each outcome.  

\psset{xunit=1cm,yunit=1cm}
\begin{pspicture}(0,0)(16,8)

\put(6.8,7){$\mathbf{C}^{\mathit{B}} = \mathit{Im}^{-1}(\lbrace \mathcal{S}^{\mathit{B}}_{\mathit{k}} \rbrace)$}
\put(1,6){$Im^{-1}(\lbrace \mathcal{A}^{\mathit{B}}_{\mathit{g}}, \mathcal{A}^{\mathit{B}}_{\mathit{e}} \rbrace)$}
\put(13.9,6){$\mathbf{O}$}
\put(4.2,6.1){\vector(1,0){9.5}}
\put(7,6.1){\line(0,-1){.7}}
\put(7,5.4){\line(1,0){.3}}
\put(7.3,5.4){\line(-2,-1){.6}}
\put(6.7,5.1){\line(2,-1){.6}}
\put(7.3,4.8){\line(-2,1){.6}}
\put(6.7,4.5){\line(2,1){.6}}
\put(7.3,4.2){\line(-2,1){.6}}
\put(6.7,3.9){\line(2,1){.6}}
\put(6.7,3.9){\line(1,0){.3}}
\put(7,3.9){\line(0,-1){.4}}
\put(6,4.7){$R$}
\put(10.8,4.7){$C$}
\put(7,3.5){\line(1,0){3}}
\put(8.5,3.5){\line(0,-1){.5}}
\put(8.3,2.8){\line(1,1){.4}}

\put(10,6.1){\line(0,-1){1.1}}
\put(9.5,5){\line(1,0){1}}
\put(9.5,4.7){\line(1,0){1}}
\put(10,4.7){\line(0,-1){1.2}}
\put(9.6,4.5){\vector(1,1){.8}}

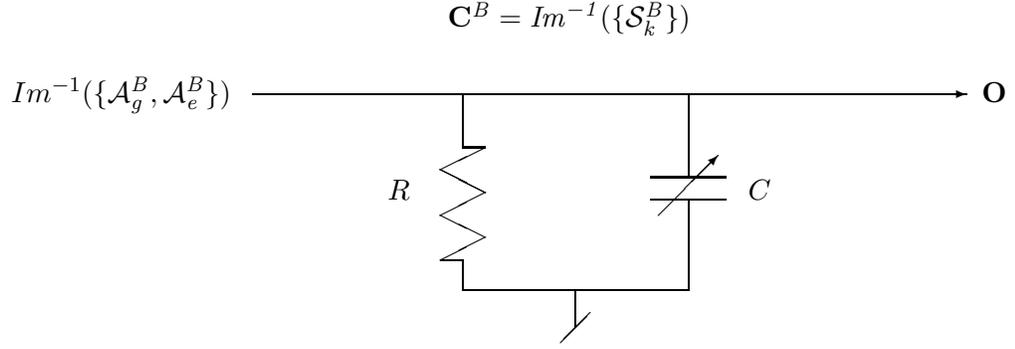
\put(.6,1){\textit{Figure 7: Simple circuit model of decoherence in an information channel $Im^{-1}(\lbrace \mathcal{S}^{\mathit{B}}_{\mathit{k}} \rbrace)$.}}
\end{pspicture}

The ``capacitance'' $C$ in Fig. 7 is clearly a measure of the ``quantum-ness'' of the channel; as $C \rightarrow 0$, the channel appears classical.  The condition $C = 0$ corresponds to infinite temporal resolution for measurement events; hence it corresponds to the ``free information'' (i.e. $\hbar \rightarrow 0$) assumption of classical physics discussed at the end of Sect. 4.  If $C = 0$, the channel stores no information about previous outcomes, so all pairs of POVMs, including those that jointly measure action commute.  The ``resistance'' $R$ measures the leakiness of the channel in either direction; as $R \rightarrow 0$, the channel approaches infinite decoherence time, i.e. perfect isolation, in the quantum ($C > 0$) case, and the ideal of noise-free communication in the classical ($C = 0$) case.

Given the representation of an information channel as an $RC$ circuit, consider a random sequence of measurements with the POVM $\lbrace \mathcal{A}^{\mathit{B}}_{\mathit{g}}, \mathcal{A}^{\mathit{B}}_{\mathit{e}} \rbrace$.  These measurements correspond to a random sequence of ``states'' of $Im^{-1}(\lbrace \mathcal{A}^{\mathit{B}}_{\mathit{g}}, \mathcal{A}^{\mathit{B}}_{\mathit{e}} \rbrace)$.  The no-cloning theorem requires that these ``states'' be non-identical, and hence that the collections of fine grained states $|\mathbf{C}(\mathit{t})\rangle$ that physically implement them be non-identical.  The individual measurement outcomes cannot, therefore, be ``remembered'' at $C$ as identical; the ``memory traces'' of distinct $|\mathbf{C}(\mathit{t_{\mathrm{1}}})\rangle$ and $|\mathbf{C}(\mathit{t_{\mathrm{2}}})\rangle$ stored at $C$ must interfere.  From $\mathbf{O}$'s perspective, this interference can be represented formally by adding a random phase factor $e^{-i \phi}$ to each transmission through the channel.  Without such interference, the signal at $\mathbf{O}$ would increase monotonically with time if measurements were made with a time separation less that $RC$, since $C$ would never fully discharge.  Such arbitrarily temporally-delocalized outcomes are never observed in practical experiments. Adding the random phase term assures that, for $t \gg RC$, interference between measurements drives the time-averaged signal at $\mathbf{O}$ toward zero.  In this purely informational $RC$-circuit model of decoherence, therefore, no-cloning is what requires the use of a complex Hilbert space to represent ``states'' in the inverse image $Im^{-1}(\lbrace \mathcal{A}^{\mathit{i}}_{\mathit{jk}} \rbrace)$ of any observable associated with an identified system.  Treating the $Im^{-1}(a^{\mathit{i}}_{\mathit{jk}})$ as names of coarse-grained basis vectors for the ``system'' $Im^{-1}(\lbrace \mathcal{S}^{\mathit{i}}_{\mathit{k}} \rbrace)$ as discussed above, an unknown quantum state of $Im^{-1}(\lbrace \mathcal{S}^{\mathit{i}}_{\mathit{k}} \rbrace)$ as measured at a future time $t$ using the $j^{th}$ available POVM $\mathcal{A}^{\mathit{i}}_{\mathit{jk}}$ can be written $|\psi^{\mathit{i}}_{j} (t)\rangle = \sum_{\mathit{k}} \alpha_{\mathit{k}} e^{-i \phi_{\mathit{k}}} |Im^{-1}(a^{\mathit{i}}_{\mathit{jk}})\rangle$ with $\alpha_{\mathit{k}}$ real, exactly as expected within standard quantum theory.

A ``quantum channel'' defined solely by non-commutativity between observables jointly measuring action is, therefore, a quantum channel as defined by standard quantum theory, provided that information is physical and the observer is a minimal observer as defined in Sect. 3.  If determinism, time-symmetry, counterfactual definiteness and decompositional equivalence are assumed, observations made through such channels satisfy the Kochen-Specker, Bell, and no-cloning theorems.  The Born rule emerges as a consequence of decompositional equivalence.  Complex phases are required by objective ignorance of the physical states implementing the channel, i.e. by no cloning.  Decoherence is understandable not as a physical process acting on quantum states, but as an intrinsic hysteresis in quantum information channels.  Measurement, in this framework, is unproblematic; \textit{if} minimal observers exist, the determinate, ``classical'' nature of their observations follows straightforwardly from their structure as classical virtual machines and the physics of a quantum channel.  The fundamental interpretative assumptions that must be added to quantum theory appear, then, to be that information is physical and that minimal observers exist.

\section{Adding Minimal Observers to the Interpretation of Quantum Theory} 

If Galilean observers are replaced by minimal observers as defined in Sect. 3, the interpretation of quantum theory is radically simplified.  The traditional problems of why some measurement bases, such as position, are ``preferred'' and how superpositions can ``collapse'' onto determinate eigenstates of those bases are immediately resolved: a minimal observer ``prefers'' the bases in which she encodes POVMs, and is only capable of recording eigenvalues in these bases.  The problem of the ``emergence'' of the classical world also vanishes: the classical world is the world of recorded observations made by minimal observers.  Minimal observers are virtual machines implemented by physical degrees of freedom; hence the classical world is a virtual world.  What the current framework adds to previous proposals along these lines (e.g. \cite{whitworth10}) is a precisely formulated model theory: the model theory expressed by the POVMs implemented by the minimal observer.

From an ontological perspective, the current framework can be viewed as an interpolation between two interpretative approaches generally regarded as diametrical opposites: a ``pure'' relative-state interpretation such as that of Tegmark \cite{tegmark10} and the quantum Bayesianism (``QBism'') of Fuchs \cite{fuchs10}.  Like QBism, the current framework views quantum states as observer-specific virtual entities.  However, instead of ``beliefs'' as they are in QBism, these virtual entities are inverse images of observer-specific POVMs in the space of possible states of the real physical world.  Like a pure relative-state interpretation, the current framework postulates a deterministic, time-symmetric Hamiltonian satisfying counterfactual definiteness and decompositional equivalence.  However, ``branching'' into arbitrarily many dynamically-decoupled simultaneous actualities is replaced by the classical notion that a sufficiently complex physical system can be interpreted as implementing arbitrarily many semantically-independent virtual machines.  Like QBism, the current framework rejects the interpretation of decoherence as a physical mechanism that generates actuality; unlike QBism, it views the ``classical world'' as entirely virtual and rejects the observer-independent ``real existence'' of bounded, separable macroscopic objects.  Like a pure relative-state interpretation, the current framework embraces non-locality as an intrinsic feature of the universe; unlike a pure relative-state interpretation, it views non-locality as a temporal relationship between instances of observation, not as a spatial relationship between objects.  The current framework is, therefore, ontologically very spare.  It postulates as ``real'' only the in-principle individually unobservable physical degrees of freedom that implement both channel and observer.  The virtual machines that are postulated are not in any sense physical; unlike Everett branches \cite{tegmark10}, there is no sense in which virtual machines constitute parallel physical actualities.  This strongly Kantian ontology is similar to that of the recent ``possibilist'' extension \cite{kastner10a} of the transactional interpretation \cite{cramer86, cramer88}, but without the notion that transactions ``actualize'' quantum phenomena in an observer-independent way. 

What the current interpretative framework emphatically rejects is the notion that the ``environment'' is a \textit{witness} that monitors quantum states and defines systems \textit{for} observers.  The idea that the environment \textit{preferentially} encodes certain ``objective'' quantum states and makes information about these states and not others available to observers is the foundation of quantum Darwinism \cite{zurek03rev, zurek04, zurek05, zurek06, zurek07grand, zurek09rev}.  It is implicit, however, in all interpretative approaches in which the classical world ``emerges'' from the dynamics in an observer-independent way.  The bounded and separable ``real existences'' postulated by QBism \cite{fuchs10}, for example, are effectively the observations of the ``rest of the universe'' viewed as an observing agent \cite{fields11b}.  The ``witness'' assumption can be found in interpretative approaches as distant in terms of fundamental assumptions from both QBism and quantum Darwinism as the possibilist transactional interpretation, where an ``experimental apparatus seems persistent in virtue of the highly probable and frequent transactions comprising it'' (\cite{kastner10b} p. 8) not from the perspective of an observer, but from the perspective of an observer-free universe.  It is this assumption of emergence via environmental witnessing that enables, explicitly or otherwise, the traditional and ubiquitous assumption of information-free Galilean observers, mere points of view or (as ``preparers'' of physical systems) points of manipulation of a pre-defined objective reality.

As pointed out in the Introduction, the logical coherence of Galilean observers must be rejected on the basis of classical automata theory alone \cite{moore56, ashby56}.  It is useful, however, to examine the Galilean observer from the perspective of the ``environment as witness.''  Consider the classic Wigner's friend scenario \cite{wigner61}, but with an omniscient ``friend'' who monitors not just an atomic decay but the states of all possible ``systems'' in the universe.  An observer can then obtain information about the state of any system by asking his friend, i.e. by interacting with the local environment as envisaged by quantum Darwinism.  A minimal observer asks his friend \textit{in language}, by executing a POVM.  The information that such an observer can obtain from the environment, whether viewed as a communication channel or as an omniscient oracle, is limited by the observer's repertoire of POVMs; a minimal observer can obtain no information about a system he cannot describe, and cannot ``observe'' that a system is in a state he cannot represent and record.  A Galilean observer, in contrast, stores no prior information and hence has no language.  Having an omniscient friend does not help a Galilean observer; they have no way to communicate.  The assumption that a Galilean observer can nonetheless obtain any information encoded by the environment is, effectively, the assumption that the observer has the same encoding capacity as the environment: what is ``given'' to the omniscient environment is also ``given'' to the Galilean observer.  This assumption was encountered at the end of Sect. 3; it is the familiar, classical assumption that information is free.

Replacing Galilean observers with minimal observers replaces the intractable philosophical problem of why observers never observe superpositions - a pseudoproblem that results from the informationally-impoverished and hence unconstrained nature of the Galilean observer - with two straightforwardly scientific problems.  The first is a problem in quantum computer science: what \textit{classical} virtual machines can be implemented by a given \textit{quantum} computer, e.g. by a given Hamiltonian oracle \cite{farhi96}?  One answer to this question is known: a quantum Turing machine \cite{deutsch85} can implement any classical virtual machine.  A second, more practical, answer is partially known: the quantum systems, whatever they are, that implement our everyday classical computers are Turing equivalent.  What we do not know is how to describe these familiar systems quantum mechanically, or how to approach the analysis of an arbitrary quantum system capable of implementing some limited set of classical virtual machines.  The second problem straddles the border between machine intelligence and biopsychology.  It is the question of what physically-realized virtual machines share POVMs, and of how these systems came to share them.  If we are to understand how multiple observers can reach an agreement that they are observing the same properties of the same thing, it is this question that we must be able to answer.

\section{Conclusion}

This paper has investigated the consequences of replacing the Galilean observer traditionally employed in interpretations of quantum theory with an observer that fully satisfies the requirements of classical automata theory.  It has shown that if both the observer and the information channel with which it interacts are implemented by physical degrees of freedom, the state space of which admits a linear measure enabling the definition of POVMs, and if the temporal dynamics of these physical degrees of freedom are deterministic, time symmetric, and satisfy decompositional equivalence and counterfactual definiteness, then the observations made by the observer are correctly described by standard quantum theory.  Quantum theory does not, therefore, require more than these assumptions.  The unmotivated and \textit{ad hoc} nature of the formal postulates that have been employed to axiomitize quantum theory, both traditionally \cite{vonNeumann32} and more recently (e.g. \cite{bub04, rau11}) can be seen as a side-effect of the assumption of Galilean observers and the compensatory, generally tacit assumption of ``axiom(o).''

The introduction of information-rich minimal observers into quantum theory brings to the fore the distinction between Shannon or von Neumann information defined solely by the dynamics and pragmatic information defined relative to an emulation mapping that specifies a control structure and hence a virtual machine.  A deterministic, time-symmetric Hamiltonian conserves fine-grained dynamic information; the von Neumann entropy of the channel $\mathbf{C}$ is zero.  Nonetheless, the pragmatic information - the list of observational outcomes - recorded by a minimal observer with an approximately ideal memory increases monotonically with time.  Pragmatic information appears, therefore, not to be conserved; ``history'' appears actual, objective and given.  This apparent asymmetry is, however, illusory.  Pragmatic information is only definable relative to an emulation mapping, a semantic interpretation of $\mathbf{C}$.  Every classical bit encoded by a minimal observer must be computed when such an emulation mapping is specified.  Hence pragmatic information is not free; it is balanced by the computational effort required to specify emulation mappings.  This effort is ``expended'' by $\mathcal{H}$ as dynamic evolution unfolds; minimal observers and the outcomes that they record are the result.  ``It from bit'' is thus balanced by ``bit from it.''

\section*{Acknowledgement}

Thanks to Eric Dietrich, Ruth Kastner and Juan Roederer for stimulating discussions of some of the ideas presented here.  Three anonymous referees provided helpful comments on the manuscript.

\bibliographystyle{mdpi}
\makeatletter
\renewcommand\@biblabel[1]{#1. }
\makeatother

\end{document}